\documentclass[reprint, onecolumn, notitlepage, times, longbibliography, aps, nofootinbib, showpacs, superscriptaddress,prd]{revtex4-1}
\usepackage[letterpaper, margin=1in]{geometry}

\usepackage[utf8]{inputenc}
\usepackage{mathtools}

\usepackage{amsthm,enumitem,graphicx,float}
\usepackage[usenames, dvipsnames]{color}
\usepackage{hyperref}
\usepackage{url}

\graphicspath{{./img/}}
\def\bdoc{\begin{document}}
\def\edoc{\end{document}}

\newcommand{\SE}[1]{\textcolor{red}{\textsf{[xx: #1]}}}
\newcommand{\TJ}[1]{\textcolor{blue}{\textsf{[TJ: #1]}}}
\newcommand{\JC}[1]{\textcolor{cyan}{\textsf{[JC: #1]}}}
\newcommand{\PN}[1]{\textcolor{blue}{\textsf{[PN: #1]}}}
\newcommand{\newtext}[1]{\textcolor{blue}{#1}}
\newcommand{\newtexta}[1]{\textcolor{magenta}{#1}}

\newcommand{\bra}[1]{\left< #1 \right|}
\newcommand{\ket}[1]{\left| #1 \right>}
\newcommand{\expVal}[1]{\left< #1 \right>}
\newcommand{\braket}[2]{\left<#1|#2\right>}

\def\beq{\begin{equation}}
\def\eeq{\end{equation}}
\def\bea{\begin{eqnarray}}
\def\eea{\end{eqnarray}}
\def\ben{\begin{enumerate}}
\def\een{\end{enumerate}}
\def\nn{\nonumber}
\def\la{\langle}
\def\ra{\rangle}
\def\a{\alpha}
\def\b{\beta}
\def\g{\gamma}\def\G{\Gamma}
\def\d{\delta}\def\D{\Delta}
\def\e{\epsilon}

\def\th{\theta}
\def\k{\kappa}
\def\l{\lambda}
\def\m{\mu}
\def\n{\nu}
\def\o{\omega}
\def\O{\Omega}
\def\p{\pi}
\def\r{\rho}
\def\s{\sigma}
\def\S{\Sigma}
\def\t{\tau}
\def\L{{\cal L}}
\def\S{\Sigma }
\def\gsim{\; \raisebox{-.8ex}{$\stackrel{\textstyle >}{\sim}$}\;}
\def\lsim{\; \raisebox{-.8ex}{$\stackrel{\textstyle <}{\sim}$}\;}
\def\gtrsim{\gsim}
\def\lessim{\lsim}
\def\loc{{\rm local}}
\def\vm{v_{\rm max}}
\def\bh{\bar{h}}
\def\del{\partial}
\def\nab{\nabla}
\def\half{{\textstyle{\frac{1}{2}}}}
\def\fourth{{\textstyle{\frac{1}{4}}}}
\def\third{{\textstyle{\frac{1}{3}}}}
\def\w{\wedge}
\def\e{\epsilon}
\def\vphi{\varphi}

\def\A{{\cal A}}
\def\bA{{\bf A}}
\def\bD{{\bf D}}
\def\bH{{\bf H}}
\def\bM{{\bf M}}
\def\bN{{\bf N}}
\def\bE{{\bf E}}
\def\bF{{\bf F}}
\def\bB{{\bf B}}
\def\bP{{\bf P}}
\def\bJ{{\bf J}}
\def\bK{{\bf K}}
\def\bL{{\bf L}}
\def\bR{{\bf R}}
\def\bS{{\bf S}}
\def\bV{{\bf v}}
\def\bv{{\bf v}}
\def\bx{{\bf x}}
\def\by{{\bf y}}
\def\bz{{\bf z}}
\def\ba{{\bf a}}
\def\bd{{\bf d}}
\def\bs{{\bf s}}
\def\bn{{\bf n}}
\def\bm{{\bf m}}
\def\bp{{\bf p}}
\def\bk{{\bf k}}
\def\bg{{\bf g}}

\def\br{{\bf r}}
\def\bnab{{\bf \nab}}

\def\bitP{\boldsymbol{P}}
\def\bphi{\boldsymbol{\phi}}
\def\btau{\boldsymbol{\tau}}
\def\bo{\boldsymbol{\omega}}
\def\bO{\boldsymbol{\Omega}}

\def\tb{\tilde{\b}}

\def\qd{\dot{q}}

\def\K1{K}

\begin{document}

%\preprint{UTTG--14--17}

%\title{Remarks on complexity and volume}
\title{Holographic Complexity and Volume}

\author{Josiah Couch}
\affiliation{Theory Group, Department of Physics, The University of Texas at Austin,
Austin, TX 78712, USA}
\author{Stefan Eccles}
\affiliation{Theory Group, Department of Physics, The University of Texas at Austin,
Austin, TX 78712, USA}
\author{Ted Jacobson}
\affiliation{Maryland Center for Fundamental Physics and Department of Physics, University of Maryland, College Park, MD 20742, USA}
\author{Phuc Nguyen}
\affiliation{Maryland Center for Fundamental Physics and Department of Physics, University of Maryland, College Park, MD 20742, USA}

\begin{abstract}
The previously proposed ``Complexity=Volume'' or CV-duality is probed and developed in several directions.
We show that the apparent lack of universality for large and small black holes
is removed if the volume is measured in units of the maximal time 
from the horizon to the ``final slice" (times Planck area). This also
works for spinning black holes. We make use of the conserved ``volume current," 
associated with a foliation of spacetime by maximal volume 
slices, whose flux measures their volume. 
This flux picture suggests that there is a transfer of the complexity from the UV to the IR in holographic CFTs, 
which is reminiscent of thermalization behavior deduced using holography. It also 
naturally gives a second law for the complexity when applied at a black hole horizon. 
We further establish a result supporting the conjecture that 
a boundary foliation determines a bulk maximal foliation without gaps,
establish a global inequality on maximal volumes that can be used to deduce the monotonicity of the
complexification rate on a boost-invariant background, and probe CV duality in the settings of 
multiple quenches, spinning black holes, and Rindler-AdS.
\end{abstract}

\maketitle

\tableofcontents

\section{Introduction}\label{Sec:Intro}

The AdS/CFT correspondence \cite{Maldacena:1997re,Witten:1998qj,Gubser:1998bc} provides a satisfying duality between a black hole in asymptotically anti de-Sitter spacetime and a thermal state of a CFT, in which the entropy of the black hole is dual to ordinary thermal entropy. What remains obscure, however, is the relation between the black hole interior and the physics of the CFT. Aside from its (semiclassical) causal isolation, the interior has two qualitative features that one would like to understand from the viewpoint of the CFT: the curvature singularity, and the growth of space. The latter follows from the well known peculiar fact that the symmetry of time flow outside the horizon becomes a space translation symmetry inside. 
Under this symmetry flow, exterior time elapses, and the length of a spacelike curve at a fixed interior radius grows, with a rate that increases without bound as the fixed radius approaches the singularity.
Susskind observed \cite{Susskind:2014rva,Susskind:2014moa,Stanford:2014jda,Brown:2015bva,Brown:2015lvg} that this growth should be reflected somehow in the CFT, because it can be captured by a gauge invariant observable of the bulk gravity theory. He proposed that it corresponds to the computational complexity of the state of the CFT, which continues to grow, after statistical equilibrium is reached, for a time that is exponential in the entropy.

This proposal was quickly refined to  
``CV-duality", according to which the complexity at a given boundary/CFT time is proportional to the volume of the maximal
slice enclosed within the  corresponding ``Wheeler-DeWitt patch,"
i.e., within the domain of dependence of a spacelike bulk hypersurface that asymptotes to the given boundary time slice\footnote{In this paper, we systematically use the term ``slice'' to refer to spacelike codimension-1 submanifolds.}. Shortly thereafter, the alternative postulate of ``CA-duality" was introduced, according to which complexity is equal to the action of the Wheeler-DeWitt patch (see \cite{Carmi:2017jqz, Carmi:2016wjl, Chapman:2016hwi, Jefferson:2017sdb,Couch:2016exn} for a selection of recent work on these two proposals \footnote{See also \cite{Harlow:2013tf} for the consideration of complexity in a related but somewhat different context involving black holes.}). Both of these proposals predict a rate of growth of the complexity at late time that roughly agrees with general expectations. 

There is reason to expect that the rate of complexification for a CFT in equilibrium scales as 
$TS/\hbar$, the product of the temperature $T$ and the entropy $S$ \cite{Susskind:2014rva}. The entropy counts the number of `active' degrees of freedom, and $\hbar/T$ is the 
timescale for thermal fluctuations. If each such fluctuation counts as the execution of a quantum gate
on active degrees of freedom, then the number of gates executed per unit time is $\sim TS/\hbar$, which is 
thus the rate at which the complexity of the state increases. In order to match this rate, the complexity for black holes that are large compared to the 
AdS radius $\ell$ should be given in terms of the volume of the maximal slice by  \cite{Susskind:2014moa}
\begin{equation}\label{CV}
    \mathcal{C} \sim \frac{V}{\hbar G\ell}.
\end{equation}
In equilibrium, the maximal slice approaches a final maximal
cylinder inside the horizon, with fixed cross-sectional area and a 
proper length that grows in proportion to Killing time. 
The above formula equates the complexity to this
area, measured in Planck units, times the proper length of the cylinder, measured in AdS length units.
For black holes small compared $\ell$, the complexity should instead be given by
\begin{equation}\label{CVr+}
    \mathcal{C} \sim \frac{V}{\hbar Gr_{+}},
\end{equation}
so that the proper length of the cylinder is measured in horizon radius units $r_+$  \cite{Susskind:2014moa}. 
Unlike the case for large black holes, this depends upon the black hole size. 
This discrepancy is a principal reason for
preferring CA over CV. The fact that the volume divisor in CV is $\ell$ for large black holes, but $r_+$ for small black holes,
indicates an apparent lack of universality.

However, in both cases this divisor actually corresponds to an intrinsic property
of the black hole: the
maximum time $\t_f$ to fall from the horizon to the final maximal cylinder is $\sim r_+/c$ for 
spherical black holes with $r_+\le \ell$ in $D\ge4$ dimensions, 
and $\sim \ell/c$ for black holes with $r_+\ge \ell$.\footnote{An alternative
but related divisor would be the proper time from horizon to final  
slice along along the volume flow. 
This scales the same way with respect to the black hole parameters as the longest time, but the numerical factor 
differs.} 
Hence the complexity formuale (\ref{CV}) and (\ref{CVr+}) actually coincide,
up to an order unity numerical factor, if the length in the denominator is understood as $d_f:=c\t_f$. That is, 
in computing the complexity, the length of a section of the final cylinder $\D L_f$ should be measured in units of the maximal 
time to fall from the horizon to the cylinder. The universal expression for the late time complexity is thus
\begin{equation}\label{CVtauf}
    \mathcal{C} \sim \frac{V}{\hbar G\tau_f}= \frac{A_f}{\hbar G} \frac{\D L_f}{\t_f},
\end{equation}
where $A_f$ is the cross-sectional area of the final slice. It turns out that $A_f$ equal to the horizon 
area $A_H$ up to an order unity factor, so that $A_f/\hbar G\sim S_{\rm BH}$ can be identified with the black hole
entropy, which is dual to the CFT entropy. The remaining factor in (\ref{CVtauf}) is then $\D L_f/\t_f$. 
In Sec. \ref{reason} we will show that, quite generally, 
\beq
\frac{\D L_f}{\t_f}\sim \kappa \D t\sim \frac{T_H}{\hbar}\D t,
\eeq
where $\k$ is the surface gravity, $\D t$ is the elapsed Killing time, and $T_H$ is the Hawking temperature of the 
black hole, which is dual to the CFT temperature. 
With these results, the expression (\ref{CVtauf}) for complexity thus becomes
\begin{equation}\label{CVH}
    \mathcal{C} \sim \frac{T_{\rm H}S_{\rm BH}}{\hbar}\D t,
\end{equation}
yielding the black hole dual of the expected complexification rate.

While the universality of the divisor $\t_f$ is more satisfying than the previous ad hoc prescription, it should be admitted that we have no rationale for measuring the length of the final slice in
units of $\t_f$, other than that gives the desired result.
Another potential drawback is that this prescription only applies to defining the complexity when the state at late times is thermal equilibrium, so that a `final' maximal slice exists. In a general dynamical setting,
this prescription is inapplicable (although as discussed in Sec.~\ref{Rindler} it {\it can} be applied in empty 
AdS, using the boost Killing field to define the notion of equilibrium).
 That said, as  discussed in the next section, the notion of complexity itself is
more ambiguous outside of a thermal setting, so it is not clear whether we should expect it to admit 
a universal holographic definition.

The CV proposal thus remains interesting, as it passes the same checks as does the CA proposal, 
in some cases
(regarding monotonicity on a stationary background) even better as discussed in Sec.~\ref{mono}. 
The purpose of this paper is to take a closer look at various aspects
of the CV proposal, attempting to sharpen it and offer some interpretation of its definition and properties, as well as to extend the tests of it. 

For both conceptual and computational reasons, we shall make 
use of a {\it volume current}, whose flux through the bulk maximal 
slices anchored at a boundary foliation is equal to the volume of those 
slices. This volume current is a unit, timelike, divergence-free vector field orthogonal to the bulk maximal foliation. 
Our interest in the role of 
this current was inspired by recent work of Headrick and Hubeny (HH) \cite{Headrick:2017ucz},
which established a ``min-flow/max-cut" theorem relating volumes of maximal slices to minimal 
fluxes of timelike, divergence-free, vector fields with norm bounded below by unity (``HH flows"). 
In Ref.~\cite{Headrick:2017ucz}, it was remarked that it is natural to relate 
minimization of the number of gates in defining the complexity of a state, in 
a dual field theory, to minimization of the flux of an HH flow in the bulk spacetime, suggesting a 
``gate-line" picture of holographic complexity. In this picture, our volume current would 
correspond to a ``gate current". 

Let us briefly describe here the HH theorem and its relation to our volume current.
The theorem states roughly that, given boundary sub-region $A$, 
the maximal spatial volume of any 
slice homologous to $A$
is equal to the minimal flux of an HH flow through the 
slice or (equivalently) through $A$.
A given minimizing HH flow has unit norm on the corresponding maximal volume 
slice, and is 
orthogonal to that 
slice, but it is not otherwise uniquely determined. By contrast, the volume current 
we employ is a particular realization of an HH flow, determined by a boundary foliation, and its flux 
gives the volume of each slice of the corresponding maximal bulk foliation. That volume is not 
conserved, because there is flux through the cutoff boundary of the bulk region. 

Although the HH theorem assumes the spacetime is orientable and time-orientable, and assumes a maximal volume slice exists, its proof does not directly invoke 
any causality assumption or energy condition on the spacetime. By contrast, the volume current requires the existence of a foliation by maximal slices. We argue in Appendix \ref{foliation} that such a foliation exists if 
i) maximal slices exist, ii) the spacetime satisfies a causality condition, and iii) the strong energy condition and Einstein equation hold. If the foliation is known to exist, then the HH theorem is a simple consequence:
the flux of any HH flow is lower-bounded by the maximal volume (for a given boundary Cauchy slice), and the theorem asserts that this bound is actually saturated. The volume current, 
when it exists, saturates this bound, so our results can be viewed as providing a 
\textit{constructive} proof of the HH theorem under certain additional assumptions.

The remainder of this paper is structured as follows.
Section II confronts the ambiguity in defining complexity. The notion seems most 
robust when applied to time evolution of thermal states, and we summarize several reasons for 
thinking the volume inside a black hole horizon captures the relevant quantity.
Section III introduces the volume current, a useful tool for quantifying properties of maximal volumes and their 
evolution, and obtains several results using it. One of these is evidence for the flow of complexity from UV to IR in holographic CFTs.
Section IV deduces a global inequality on maximal volumes, and uses this to establish
the monotonic increase of the rate of volume growth on a boost invariant background. 
Section V probes CV duality in three settings: black hole formation with one or two shells of matter, 
spinning black holes, and empty AdS viewed as a pair of thermal Rindler wedges. 
Section VI is a brief conclusion and outlook. In Appendix A it is argued, assuming
the existence of maximal 
slices, a causality condition and the strong energy condition,  that a boundary foliation determines a maximal volume bulk foliation. The unit vector field normal to this foliation is the volume current. The remaining three appendices
derive useful technical results. 
For the balance of this paper we use Planck units, with $\hbar=c=G=1$.

\section{Volume inside and outside the horizon}
\label{InsideOut}

The complexity of a pure quantum state is a measure of how many simple unitary operations, or ``gates," it takes to produce it, starting with some reference state \cite{Cleve:1999vn, 2008arXiv0804.3401W, Aaronson:2016vto}. Hence, in general, complexity is defined only relative to the choice of reference state and the choice of gates. The original motivation for the proposal of CV duality pertained to time development of complexity at the thermal scale in a finite temperature pure state. In this context, the reference state could presumably be taken to be the thermal microstate at any fixed time, and the gates could be taken to be a fixed collection of gates that act at the thermal energy and length scales, so the rate of change of complexity is intrinsically defined without significant arbitrariness.

However, the CV proposal encounters a 
divergence in asymptotically AdS spacetime, where the volume of a maximal slice diverges at spatial infinity. This divergence occurs for any state and, according to the usual UV-IR relation in AdS/CFT duality, it would presumably correspond, according to CV duality, to a divergent UV complexity of the CFT vacuum. While the vacuum is simply the ground state of the theory, it is complex if considered as a state to be prepared, starting with a spatially unentangled state, by the application of local quantum gates. Some analysis has suggested that this interpretation of the UV limit of CV duality might be sensible \cite{Jefferson:2017sdb},  
although the volume-complexity relation could be infinitely sensitive to the somewhat arbitrary definition of the reference state and gates, and to sub-leading modifications of the short distance structure of the state \cite{Moosa:2017yvt,Moosa:2017yiz}.

The volume divergence has generally been dealt with in the literature by imposing a cutoff at some large radius, and focusing on the time dependence of the volume, which does not depend on the location of the cutoff. This corresponds, in effect, to taking the reference state to be the vacuum above the cutoff energy scale, and some ``unentangled" state below that scale. The rate of change of the volume in a stationary,
thermal state at late times is independent of the location of the cutoff, because the volume growth all happens inside the horizon of the black hole. 
For this reason, and several others, it makes a lot of sense to count only the volume behind the horizon:\footnote{It was also noted in Ref.~\cite{Susskind:2014jwa}
that the volume divergence can be regulated by counting only the volume behind the horizon.}
\begin{itemize}
\item It is only the complexity at the thermal scale that appears to have a robust significance, independent of the arbitrary choices of reference state and gates. 

\item The complexity divisor of the volume, as explained in the introduction, is universal when recognized as a free-fall time from the horizon to the final maximal slice.

\item The stationary state 
volume growth at late time occurs behind the horizon. This was explained in a picturesque way in \cite{Susskind:2014moa},
where it was referred to as ``unspooling complexity" from the horizon. 

\item If the reference state is the vacuum, then only black hole states have complexity that scales as $O(N^2)$ in the CFT. This suggests that holographic complexity (with a vacuum reference state)
should, at leading order in $N$, be associated only with black holes, 
and that the relevant volume in CV duality should be only that located behind a horizon.

\item For a two-sided black hole, a natural reference state is the thermofield double,
which is a ``Euclidean vacuum" for this topology. 
The maximal bulk slice corresponding to this state is a global time slice invariant under time reflection 
(like the $t=0$ slice in Schwarzschild coordinates), which does not enter the (future or past) horizon, and therefore has zero volume behind the horizon. The volume behind the horizon thus gives
the ``right" result: the complexity vanishes, 
since the reference state by definition has zero complexity, but it grows if the time on
one boundary is boosted relative to that on the other.

\item A ``second law of complexity" \cite{Brown:2016wib, Brown:2017jil} follows directly when the horizon is a causal barrier, as discussed in the next section. 

\item The volume inside a white hole horizon can also contribute to the complexity, as
in the shockwave scenario discussed below. This 
allows for \textit{decreasing} complexity, when such behavior is expected.

\item In the extremal limit of rotating or charged black holes, the exterior of the horizon develops an infinitely long throat. Regularizing the volume near the boundary (or, for that matter, anywhere outside the horizon) would predict that the complexity of the thermofield double state diverges in the IR as extremality is approached. This questionable feature is avoided by regularizing at horizon.

\end{itemize}

When applied in a general, time dependent setting, the proposal that 
complexity corresponds only to the volume behind the horizon suffers from a major drawback, however, if we use the {\it event horizon}, because the volume inside can grow {\it before} anything changes in the CFT, at the boundary of the maximal slice. 
This is illustrated by an example in Sec. \ref{Sec:Vaidya} We therefore propose to use an ``apparent horizon" as the cutoff surface when there is time dependence. We follow the prescription, used previously in the literature, of measuring the volume on leaves of a foliation of the spacetime by spacelike hypersurfaces that maximize the volume inside an outer cutoff boundary. The apparent horizon is then defined as the boundary of the region containing trapped surfaces on each leaf of this foliation. In (quasi)stationary black hole spacetimes, this apparent horizon will (nearly) coincide with the event horizon. A component of the apparent horizon that asymptotes to the event horizon consists of points lying on marginally outer trapped surfaces \cite{Bengtsson:2010tj}.

When focusing on the volume inside the horizon, we are limited to discussing the growth of complexity in states dual to a spacetime with a horizon. This is not as restrictive as it might seem, since even empty AdS is a thermal state, when viewed as a pair of Rindler wedges. Indeed, the much studied account of complexity increase for the two-sided black hole can be adapted in a straightforward manner to the Rindler case, where the relevant volume is that inside the Rindler horizon. The interpretation in this case appears to be fully consistent with that for black holes, as we explain in Sec. \ref{Sec:FurtherProbes}.

\begin{figure}[h!]
    \centering
    \includegraphics[width=6cm]{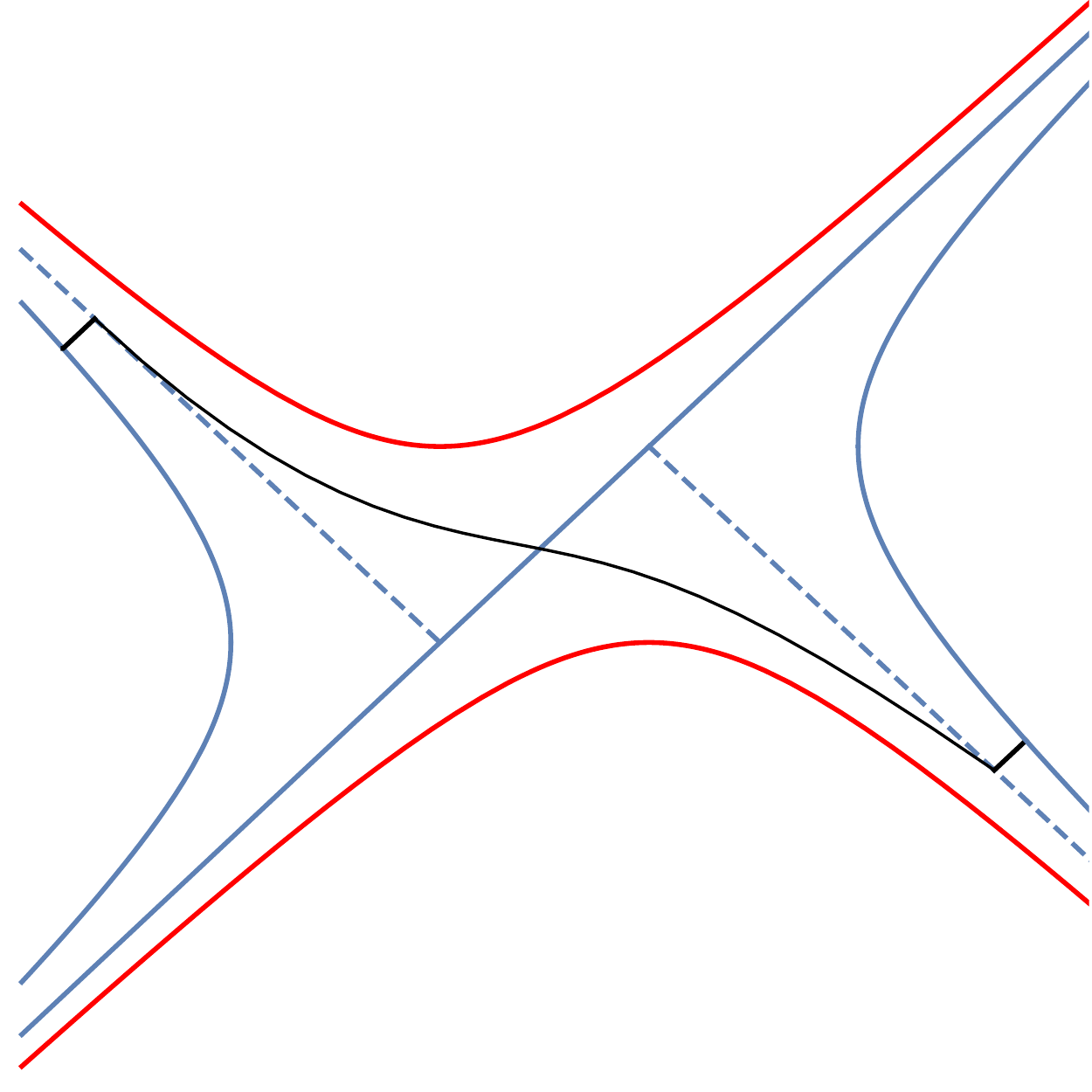}
    \caption{Shockwave geometry dual to a perturbed thermofield double state, with a maximal volume hypersurface anchored at late time on the left and early time on the right.}
    \label{fig:Switchback}
\end{figure}

So far we have been referring to the volume inside the black hole horizon, which is relevant for late time equilibrium states. However an important test for any proposed holographic dual of complexity is that it exhibit the switchback effect \cite{Stanford:2014jda}, which brings the white hole horizon into play. The switchback effect refers to a small time-deficit in the growth of complexity, of order twice 
the ``scrambling time", when a state is evolved backwards in time, perturbed relative to the reference state, and then evolved forwards in time. 
The calculations in \cite{Stanford:2014jda} demonstrated that the volume of maximal 
slices does 
holographically capture the switchback effect for the thermofield double state, and in particular the maximal volume 
slices can traverse the black hole region on one side of the shock, and the white hole region on the other side. In the late time approximation used in  \cite{Stanford:2014jda}, the portion of the maximal 
slice outside the horizons does not contribute to the total volume, because it is null, as illustrated in Fig. \ref{fig:Switchback}.
Hence the volume inside the black and white hole horizons suffices to capture the switchback effect. 
In general, therefore, our proposal must be taken to include the volume inside the white hole horizon. This appears somewhat natural, considering the fact that the derivation of the switchback effect involves reversed time evolution, and the time reverse of a black hole is a white hole.

Finally, although it appears difficult to relate the volume outside the horizon to a definition of complexity of the state in a universal manner, the  assumption that such a relation exists leads to the interesting picture of complexity flowing from UV to IR, as explained in the following section.

\section{Volume current}\label{Sec:VolumeFlow}

While the volume of maximal 
slices is a nonlocal construct, there is an associated local object, the ``volume current," which can be used to infer the volume growth behind the horizon and the second law of complexity, and which 
is suggestive the UV to IR flow of complexity. In this section we introduce the volume current, and use it 
to establish several important properties of the proposed CV duality.

A volume current will be defined given a foliation of spacetime by spacelike hypersurfaces with maximal volume. In the present application, we are interested in asymptotically AdS spacetimes, in which a maximal foliation is determined by 
a Cauchy foliation of the boundary by slices orthogonal to an asymptotic Killing flow defining time translation. Provided that there is a unique bulk maximal 
slice that terminates on any fixed boundary Cauchy slice, and provided these bulk 
slices do not skip over a ``gap" in the bulk, the boundary foliation induces a bulk foliation by maximal 
slices $\S_t$, labeled by a parameter $t$. 

\begin{figure}[h!]
    \centering
     \includegraphics[width=5cm]{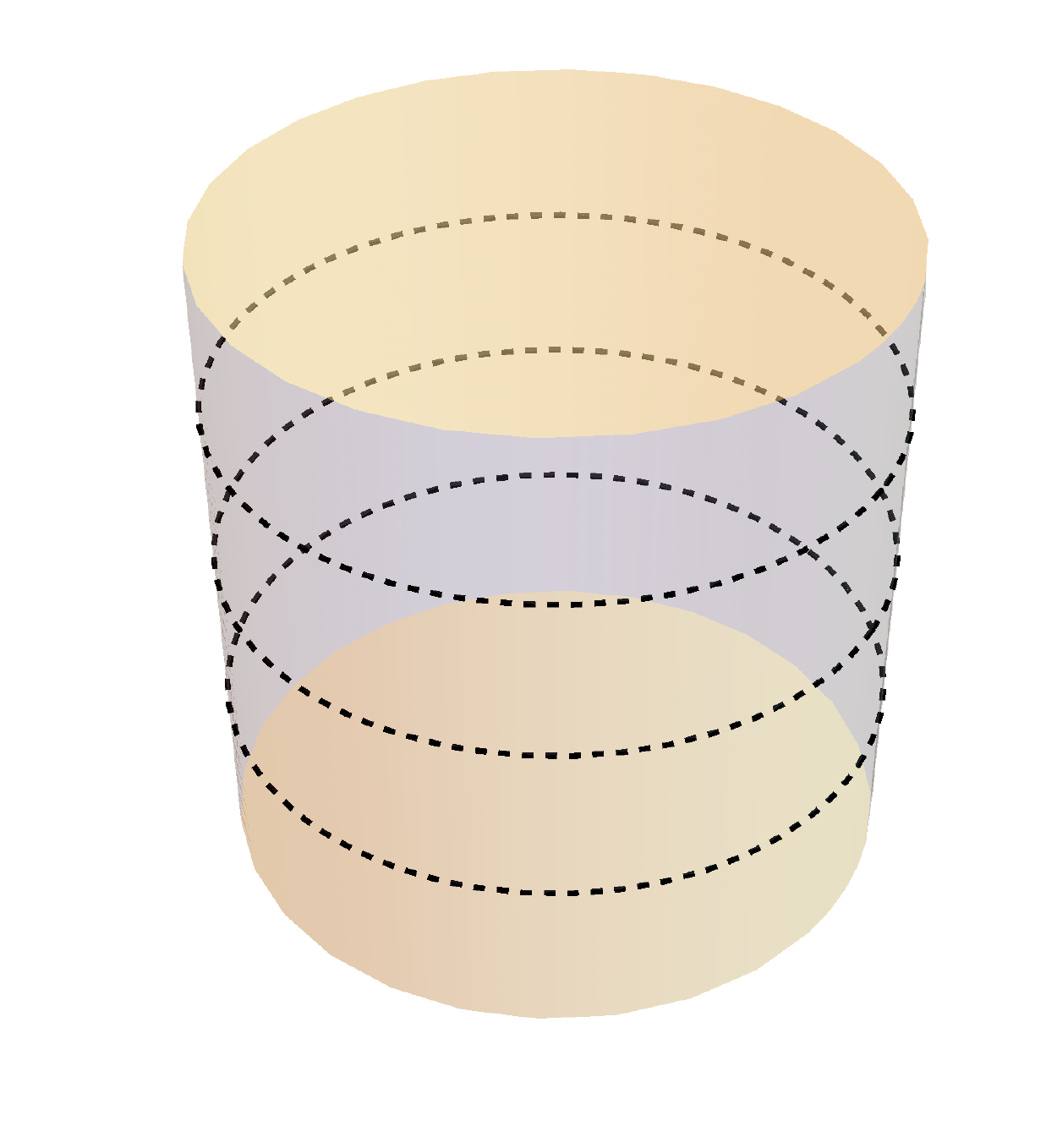}
     \includegraphics[width=5cm]{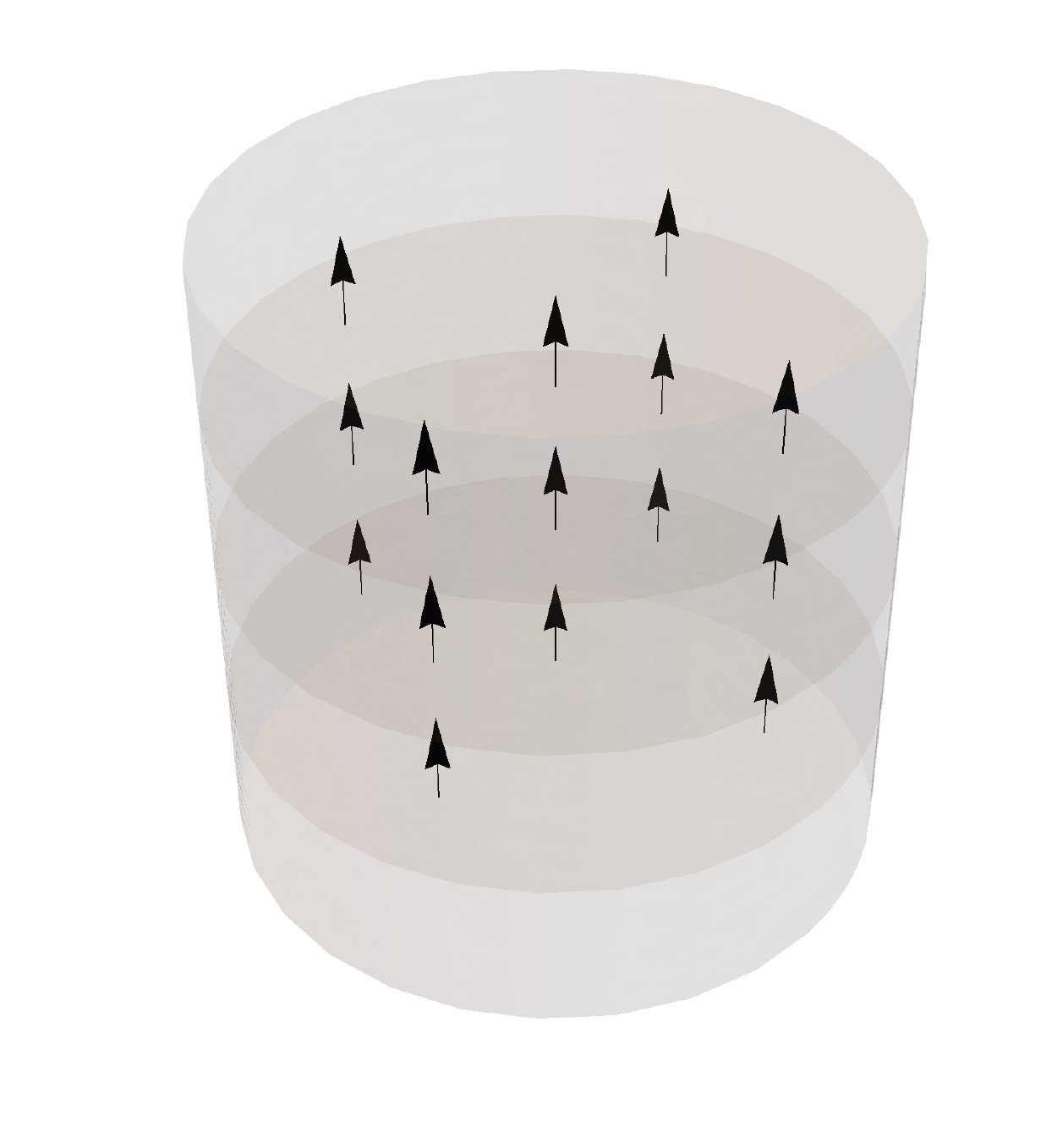}
    \caption{Left: Illustration of the boundary foliation $\Sigma{(\tau)}$ with 3 slices in the foliation. Right: Illustration of the corresponding bulk foliation by maximal 
    slices, and the volume flow.}
    \label{fig:ComplexityFlow}
\end{figure}

We establish the existence of such a bulk foliation by a reasonably convincing---if not mathematically rigorous---series of arguments in Appendix \ref{foliation}. To rule out the possibility of gaps we will need to assume that the timelike convergence condition 
(which is equivalent to the strong energy condition modulo the Einstein equation) holds. Whether or not a global foliation exists, 
our construction can be applied to the portion of spacetime prior to the final slice that is foliated without a gap.
The divergence of the unit timelike vector field $v$ orthogonal to the bulk foliation is the trace of the extrinsic curvature $K$ of $\S_t$, which vanishes since the 
slices are assumed to be maximal. This vector field is thus a conserved current, which we dub the {\it volume current} associated with the 
maximal foliation. The volume of $\S_t$ is the flux of this current through $\S_t$,  
\beq
V_t=\int_{\S_t} v\cdot\e.
\eeq
(Here $\e$ is the spacetime volume element, and the dot indicates contraction on the first index of $\e$.)  
The construction of the volume flow $v$, starting from a boundary foliation, is illustrated in Figure \ref{fig:ComplexityFlow}.

As discussed in Sec. \ref{InsideOut},
to obtain a finite volume, and hence a finite putative complexity, the integral must be cut off at some outer boundary $\del\S_t$. We will continue to use the letter ``$V$" for this truncated volume. 

\subsection{Second law of complexity}

Since the divergence of $v$ 
vanishes, the change $\D V$
from one time slice to another is entirely accounted for by the flux of $v$ through the 
boundary, or boundaries, of that slice. If we restrict to the volume inside the horizon, then the change is accounted for by the flux of $v$ through the horizon. 
Since $v$ is a future pointing timelike vector, the flux through the future event horizon is positive, and it follows that the interior volume can only increase. When considering a spacelike portion of the apparent horizon forming a past boundary of the trapped region, again the flux is positive.
CV duality then implies that in these situations, the complexity must increase, in accordance with the second law of complexity \cite{Brown:2016wib, Brown:2017jil}.
Note that this argument applies in arbitrary dynamical black hole spacetimes, such as a black hole formed by collapse. If, however, the apparent horizon has a timelike section, which can happen when a black hole evaporates, and even when positive energy conditions hold \cite{Bengtsson:2010tj}, then we cannot rule out a decrease in the volume enclosed. This seems natural: when this horizon is not a causal barrier, there is no reason to expect the associated complexity to irreversibly increase.

Note that when the region behind the horizon includes the white hole, as with the two-sided black hole with a shockwave of Figure \ref{fig:Switchback}, the complexity can {\it decrease} as time increases on
the side opposite to the shock \cite{Stanford:2014jda}. Correspondingly, the volume of the maximal 
slice inside the white hole decreases, since the volume current can only exit the white hole horizon.

\subsection{Complexity flow from UV to IR}
The flux of the volume current inward across the horizon suggests a picture of complexity 
flowing from UV to IR, which is further corroborated by examination of the flux through
other surfaces. Consider the 
section of a maximal  
slice $\S_t$ stretching between the horizon of a black hole and the outer cutoff boundary in asymptotically AdS spacetime. The rate of complexification of the 
thermal degrees of freedom should not depend upon where the cutoff surface is placed, because
that just changes the constant complexity assigned to the degrees of freedom that are in 
their ground state.  Holographically, this works because at sufficiently large radius,
as explained below,  $v$ is 
invariant under the asymptotic Killing flow. 
The volume between two large radii is thus independent of time, which
implies that the flux of $v$ through the boundary is independent of its (large) radius.
Moreover, at sufficiently late times, as also explained below, 
$v$ is invariant under the Killing flow {\it everywhere}, 
including on and inside the horizon. The flux 
of volume through the horizon is therefore equal to the flux through the outer boundary at the UV cutoff.
According to CV duality, 
{\it the complexity thus flows from the UV to the IR, and accumulates at the thermal scale}.

This conclusion may be related to the fact that, 
in a holographic CFT, thermalization proceeds 
from UV to IR \cite{Balasubramanian:2010ce,Balasubramanian:2011ur}.
On the other hand, it
seems to be somewhat in tension with the
fact that in a thermal state the UV degrees of freedom remain unexcited. If unexcited, 
how could they participate in the generation of complexity? Perhaps since their
excitation is not strictly zero, but only exponentially suppressed, their dynamics could
provide the source from which the complexity unfolds. Or is complexity generated 
purely from the thermal scale fluctuations? And if the latter is the case, 
then how can we understand the dual flow of the volume current from large to small radii?
We leave these questions to be addressed in the future.

\subsection{Asymptotic and late time volume flow on stationary spacetimes}

\begin{figure}[h!]
    \centering
    \includegraphics[width=6.5cm]{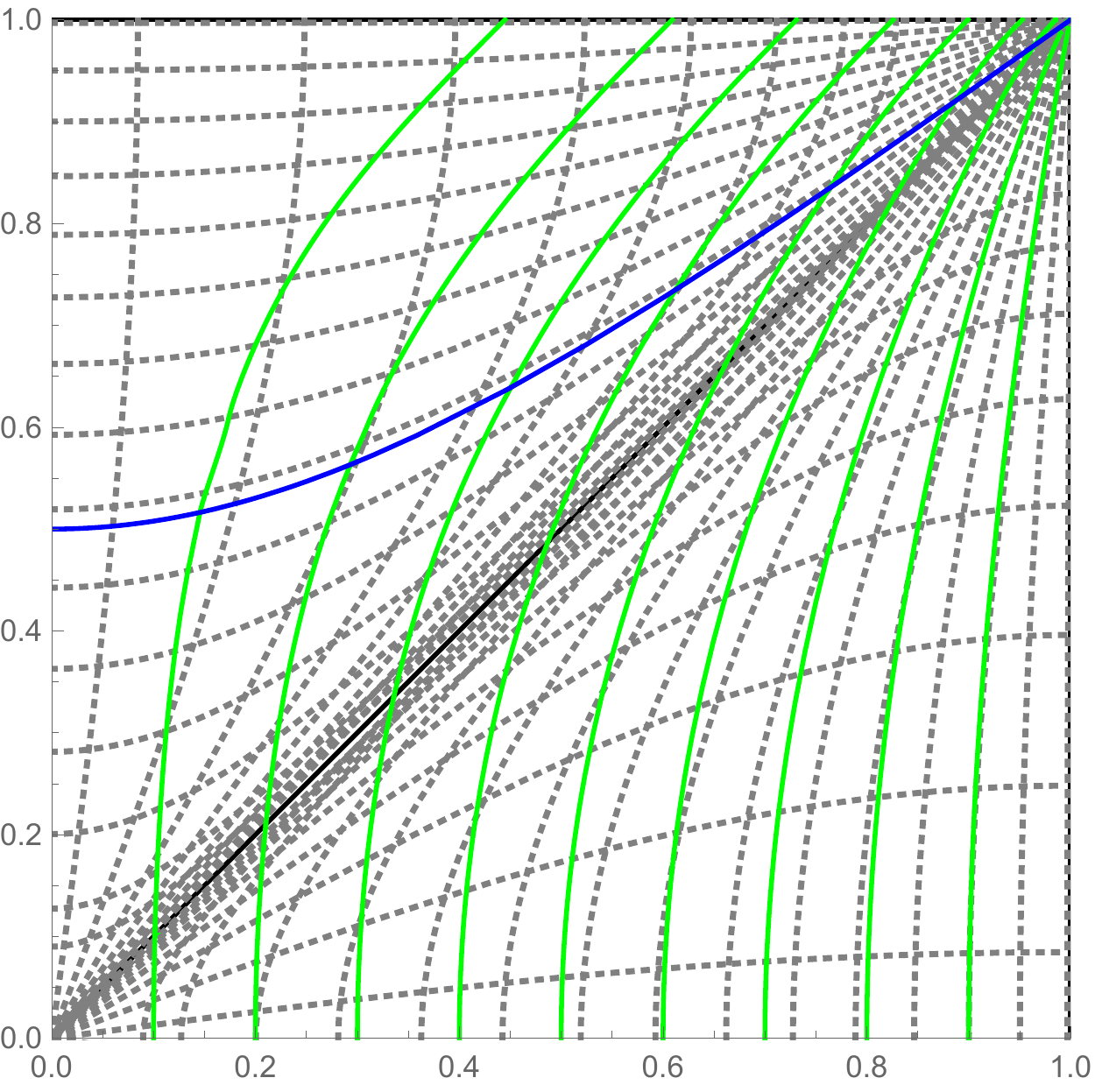}
    \includegraphics[width=6.5cm]{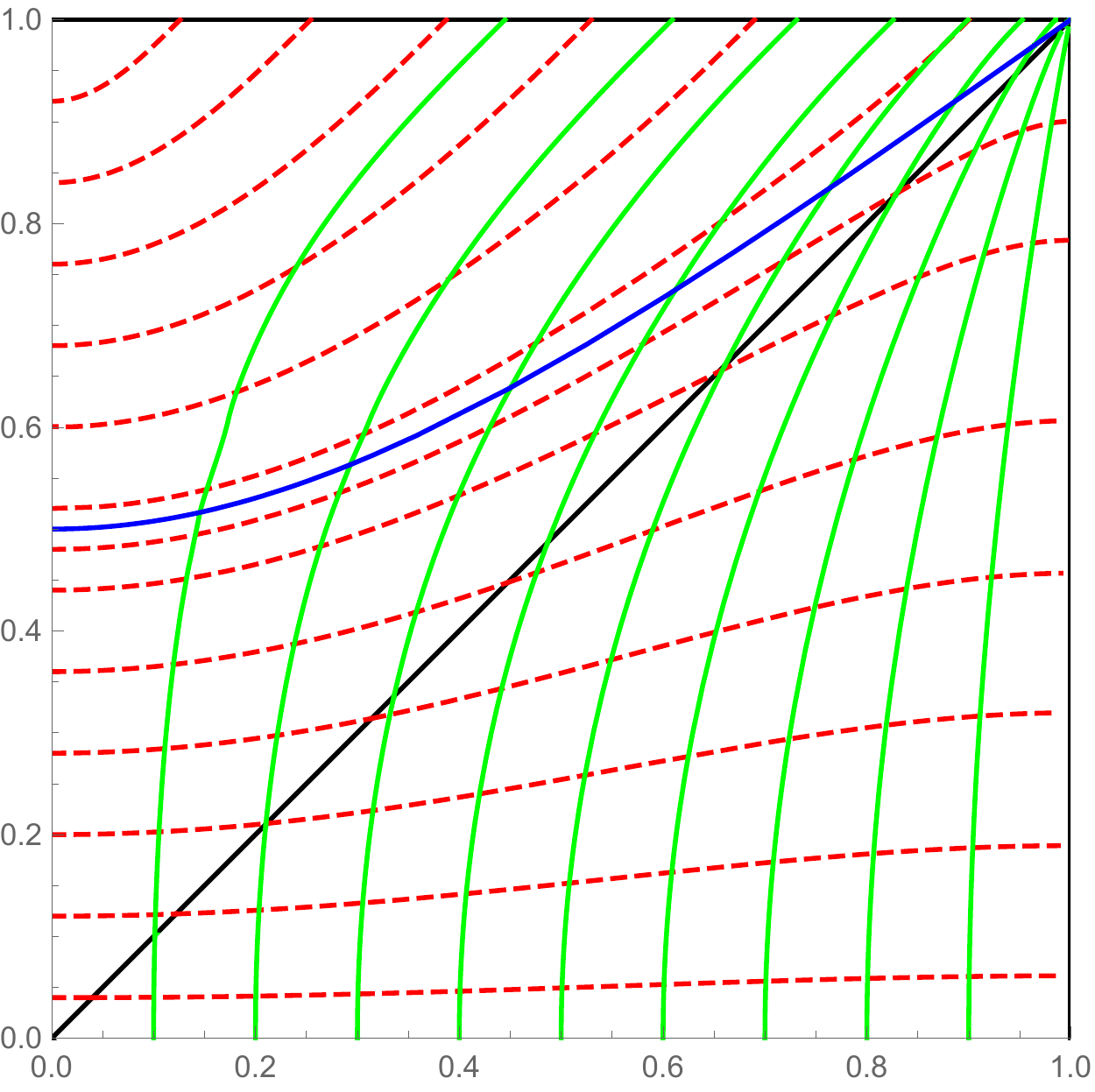}
    \caption{Plot of the flow lines of the BTZ volume current (in solid green) on a quarter of a Penrose diagram. Left: the flow lines are shown together with the Schwarzschild coordinate grid lines (dotted black), and the final slice (solid blue). Right: the flow lines are shown together with the maximal 
    slices (dashed red).}
    \label{fig:VolumeFlowDisplay}
\end{figure}

At sufficiently large radius $v$ is 
invariant under the asymptotic Killing flow, because the maximal 
slices must asymptote 
to the boundary slices defining the maximal foliation, which are taken into each other by 
the Killing flow. Nevertheless, in the two-sided eternal black hole spacetime, 
the Killing flow does not push both boundary slices to the future together, so in general the 
slices of the corresponding maximal foliation are not related by the Killing flow. 
However, at sufficiently late times the foliation becomes invariant under the Killing flow.
This can be seen from the fact that the maximal slices approach the final maximal slice
inside the horizon, and the final slice is taken into itself under the Killing flow. 
In Appendix \ref{App:FluxEquality}, we demonstrate these claims
by explicit computation with the AdS-Schwarzschild black hole. Figure \ref{fig:VolumeFlowDisplay} shows a plot of the volume current for the BTZ black hole,
based on the analytical expression derived in \ref{App:BTZFlux}.
This figure illustrates 
the asymptotic invariance under the Killing flow at the boundary and as the final slice is approached.

The late time limit of this flow can be easily found in 
closed form in spherical symmetry, where it is given by $v = v^t \partial_t + v^r \partial_r$ in 
Schwarzschild coordinates. The normalization condition $v^2=-1$
determines $v^t$ in terms of $v^r$.  At late times the components are independent of $t$, and 
the divergence free condition implies $v^r=-K/r$ where $K$ is some constant. 
$K$ can be determined by the normalization condition $g_{rr}(v^r)^2=-1$ on the 
$t=0$ line in the middle of the black hole interior region since, by symmetry, 
$v^t$ vanishes there. Because we have assumed the late time limit, this must be done 
at the ``final slice" \cite{Stanford:2014jda,Susskind:2014moa},
which is the maximal 
slice at constant $r$ in the black hole interior. 
For example, in the non-rotating 
case of the BTZ black hole treated in 
Sec. \ref{Sec:FurtherProbes} we have 
\beq\label{K}
\K1=r_f\a_f=r_+^2/2\ell.
\eeq
where $\a_f$ is the norm of the Killing vector $\partial_t$ at $r_f$.

The constant $\K1$ gives the 
rate of volume flow, with respect to Killing time,  per unit angle, 
through any surface of constant $r$ coordinate, e.g.\ the horizon. 
To see this, note that the volume flux is given by the integral of $v\cdot\e$ pulled back to the
constant $r$ surface. In $t,r,\phi$ coordinates, $\e=r\,dt\w dr\w d\phi$, so this pullback is
$-rv^r\, dt\w d\phi=\K1\,dt\w d\phi$. This conclusion generalizes to spherical black holes in any 
spacetime dimension.

\subsection{Asymptotic volume growth and complexity}
\label{reason}
For the BTZ black hole \cite{Banados:1992wn}, the $\K1$ written above can be 
expressed in terms of the surface gravity $\k=r_+/\ell^2$ and the horizon area $A=2\pi r_+$
as 
\beq\label{K1}
\K1=\ell \k A/4\pi=2\ell T_{\rm H}S_{BH},
\eeq
where $T_{\rm H}$ and $S_{BH}$ are the black hole temperature and entropy,
respectively. In this way, we can see that the late-time rate of growth is $2\ell$ times $T_{\rm H}S_{BH}$.
The factor $\ell$ is the ``divisor," discussed in the introduction, that gives the ratio of volume to complexity.

The fact that $\K1\propto \ell T_{\rm H}S_{BH}$ is not an accident. It could be anticipated from the first equality in \eqref{K}.
In fact, that equality generalizes to a $D$ dimensional spherically symmetric spacetime, where
\beq\label{K2}
\K1=r_f^{D-2}\a_f,
\eeq
and to black holes of any size.
This can be used to understand why, as mentioned in the Introduction,
the ratio of volume to complexity should be the maximal proper time from the horizon to the final 
slice for black holes of any size. 
The factor $r_f^{D-2}$ is the area per solid angle of a cross section of the final slice. It turns out that $r^{D-2}\a(r)$ reaches its maximum 
not far from the horizon, so we have $r_f\sim r_+$. The first factor 
in \eqref{K2} therefore scales as the horizon area per solid angle, times a numerical constant.
The factor $\a_f$ is the norm of the Killing vector $\partial_t$ at the final slice. The Taylor expansion for
$\a$ around the horizon is $\a=\k \t +\dots$, where $\t$ 
is the proper time from the horizon, in the direction 
orthogonal to the Killing flow.\footnote{Quite generally, 
$\k=|d\a|_{\rm horizon}$, where $\a$ is the 
norm of the horizon generating Killing vector. Usually one sees this relation applied to the gradient in the spacelike
direction from the bifurcation surface, but it can equally well be applied in the timelike direction as done here.}
Thus, for both large and small spherical black holes in any dimension, the 
volume grows at a rate
\beq\label{K3}
\K1\sim \t_f \k A_+ \sim \t_f T_{\rm H}S_{BH}
\eeq
where the symbol $\sim$ denotes equality up to numerical constant that depends on spacetime dimension and is
different for large and small black holes. Since complexity is expected to grow in the dual CFT at the rate $\sim TS$, we conclude that
the ratio of volume to complexity should be $\t_f$ (i.e. $\hbar G \t_f$). In section \ref{Kerr} we show that
this reasoning also applies to the Kerr metric (with vanishing cosmological constant).

\subsection{Maximal time from horizon to final slice}

In this subsection we first compute the maximal time $\t_f$  from the horizon to the final slice for
hyperbolic, planar, and spherical Schwarzschild-AdS black holes.
We next give a general argument, analogous to that used in Hawking's cosmological singularity theorem, 
showing that for any black hole in a spacetime with negative cosmological constant $\Lambda$,
and satisfying the strong energy condition for matter other than the cosmological constant, $|\Lambda|^{-1/2}$ sets
an upper bound for the value of $\t_f$. 

\subsubsection{Schwarzschild-AdS black holes}
The value of $\t_f$ for Schwarzschild-AdS black holes is given by the proper time from 
$r_+$ to $r_f$ along the line $t=0$:
\beq
\t_f = \int_{r_f}^{r_+} \frac{dr}{\a(r)}.
\eeq
To estimate the value of this integral, we may use the Taylor expansion about the horizon.
The line element has the form $ds^2 = -\a^2\, dt^2 + \a^{-2}\,dr^2 + r^2 h_{ij}dx^i dx^j$,
and $(d\a^2/dr)_+=(-2d\a/d\t)_+ =-2\k$, so  
\beq
\t_f \approx \frac{1}{\sqrt{2\k}}\int_{r_f}^{r_+} \frac{dr}{\sqrt{r_+-r}}= \frac{\sqrt{2(r_+-r_f)}}{\k}
\sim \frac{\ell r_+}{\sqrt{(D-1)r_+^2 + (D-3)k\ell^2}},
\eeq
where $k=-1,0,1$ for hyperbolic, planar, and spherical black holes, respectively. Thus for the BTZ black hole
($D=3$) or planar black holes, or hyperbolic or spherical black holes with $r_+\gg \ell$, we have $\t_f\sim \ell$.
If instead $r_+\ll\ell$ and $D \ne 3$ and $k \ne -1$, then $\t_f\sim r_+$.

The case of small hyperbolic black holes should be treated separately: this case has an extremal limit, i.e. a lower bound for $r_{+}$ of the order of $\ell$ \cite{Emparan:1999gf}. The estimate for $\tau_{f}$ above assumes that $r_{+} - r_{f} \sim r_{+}$ upto some order unity factor, but $r_{+}-r_{f} = 0$ at extremality. 
A computation expanding around extremality (similar to that for the Kerr case treated below in section \ref{Kerr}) shows that $\tau_{f} \sim \ell$ for the extremal hyperbolic black hole.

\subsubsection{Upper bound to $\t_f$ set by the AdS scale} 
It is interesting to note that an upper bound of the form 
$\t_f\lesssim \ell$ follows from a more general result. Consider the future
domain of dependence $D^+(S)$ of any achronal spacelike surface $S$ (not necessarily a Cauchy  
slice for the whole spacetime). 
$D^+(S)$ is itself a globally hyperbolic spacetime, so 
Theorem 9.4.5 in \cite{Wald} tells us that any point $p$ in $D^+(S)$ lies on a curve that maximizes the time to $S$, and  
Theorem 9.4.3 implies that this curve is a geodesic that meets $S$ orthogonally, without a conjugate point between $p$ and $S$. 
Integration of the Raychaudhuri equation for the congruence of geodesics orthogonal to $S$ then 
shows that, regardless of the value of the trace of the extrinsic curvature of $S$
(which is the expansion of this congruence evaluated at $S$), 
each geodesic must have a conjugate point within a time 
$\int_{-\infty}^{\infty}d\theta/(\theta^2/2+R_{ab}u^au^b)$. 
To obtain this we neglect the squared shear term in the Raychaudhuri equation, 
since it would only make the time shorter. If there is a negative cosmological constant $\Lambda=-(D-1)(D-2)/2\ell^2$, 
and if the matter stress energy tensor satisfies the strong energy condition and the Einstein equation holds, then 
by neglecting the matter contribution we obtain the upper bound 
$\pi\sqrt{D-2}/\sqrt{-\Lambda}=\sqrt{\frac{2}{D-1}}\pi\ell$ to the time.
No point inside $D^+(S)$ can lie at a time greater than this from $S$. For example, 
this result applies to the domain of dependence of a Cauchy  
slice for an asymptotically AdS spacetime,
also known as the ``Wheeler deWitt patch". 

To derive an upper bound for $\t_f$, we can apply this result to the case where the achronal surface $S$ is a spacelike 
slice just inside the future horizon $H$ of the black hole. As long as the final slice lies inside $D^+(H)$,\footnote{If $D^+(H)$ does not contain everything inside the event horizon, there is a Cauchy horizon, which is presumably unstable to formation of a singularity, eliminating the Cauchy horizon.} we obtain the upper bound 
\beq
\t_f\le \sqrt{\frac{2}{D-1}}\pi\ell.
\eeq
For $D=3$ this becomes $\t_f\le\pi\ell$, which is consistent with the exact result $\t_f=\pi\ell/4$ 
obtained in section \ref{rotating} for the rotating BTZ black hole.

\section{Global volume inequality and complexification rate monotonicity}\label{Sec:MixedSecondDerivative}
\begin{figure}[h!]
    \centering
     \includegraphics[width=7cm]{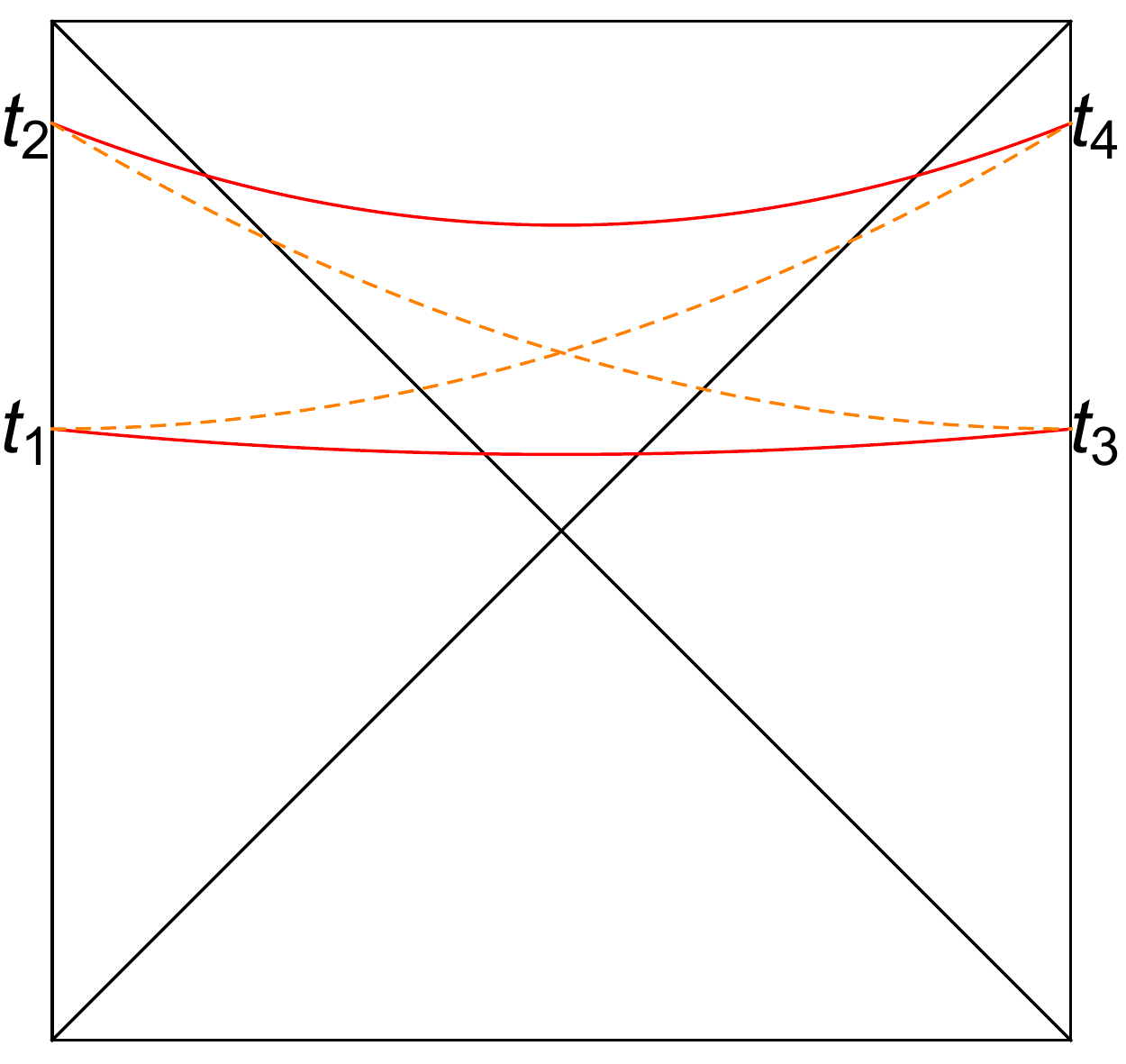}
    \caption{An SSA-like inequality is obeyed between the four maximal slices shown (solid red and dashed orange).}
    \label{fig:SSA}
\end{figure}

In this section we discuss a global inequality 
relating the volume on different slices, which 
leads to an inequality on mixed partial derivatives with respect to boundary
time. On a boost symmetric background,  this allows us to obtain an 
inequality for the second time derivative of the volume, which implies
that the complexification rate grows monotonically on boost invariant black hole backgrounds. 
We thus recover from a general viewpoint this fact found previously using explicit computations with eternal black holes in AdS.
To derive the global volume inequality, 
we need only use the definition of maximal 
slices; no energy condition or other additional ingredient is needed.
Consider for concreteness a compact box in an eternal black hole spacetime (Figure \ref{fig:SSA}), with the vertical sides of the box taken to be some near-boundary cutoff. Let $t_{1}$, $t_{2}$ be two times on the left cutoff (with $t_{1} < t_{2}$), and $t_{3}$, $t_{4}$ be two times on the right cutoff (with $t_{3} < t_{4}$). 
The inequality then says that
\begin{equation}\label{SSA1}
    \mathrm{Vol}{(t_{1},t_{3})} + \mathrm{Vol}{(t_{2},t_{4})} - \mathrm{Vol}{(t_{1},t_{4})} - \mathrm{Vol}{(t_{2},t_{3})}\geq 0,
\end{equation}
where $\mathrm{Vol}{(t_{1}, t_{3})}$ is the maximal volume between time $t_{1}$ on the left and time $t_{3}$ on the right, etc.
Note that, even though each of the four maximal 
slice volumes diverges as the cutoff is sent to the boundary, the linear combination in (\ref{SSA1}) is UV-finite. 

To establish the inequality, observe that the two dashed orange 
slices in Fig.~\ref{fig:SSA} intersect each other, and we can divide them into four segments, each connecting the intersection with one of the four boundary times. By maximality, we know that $\mathrm{Vol}{(t_{1},t_{3})}$ is greater than the sum of the volumes of the two lower orange segments. Similarly, we know that 
$\mathrm{Vol}{(t_{2},t_{4})}$ is greater than the sum of the volumes of the two upper orange segments. The sum of these two inequalities yields (\ref{SSA1}). 

\begin{figure}[h!]
    \centering
     \includegraphics[width=4cm]{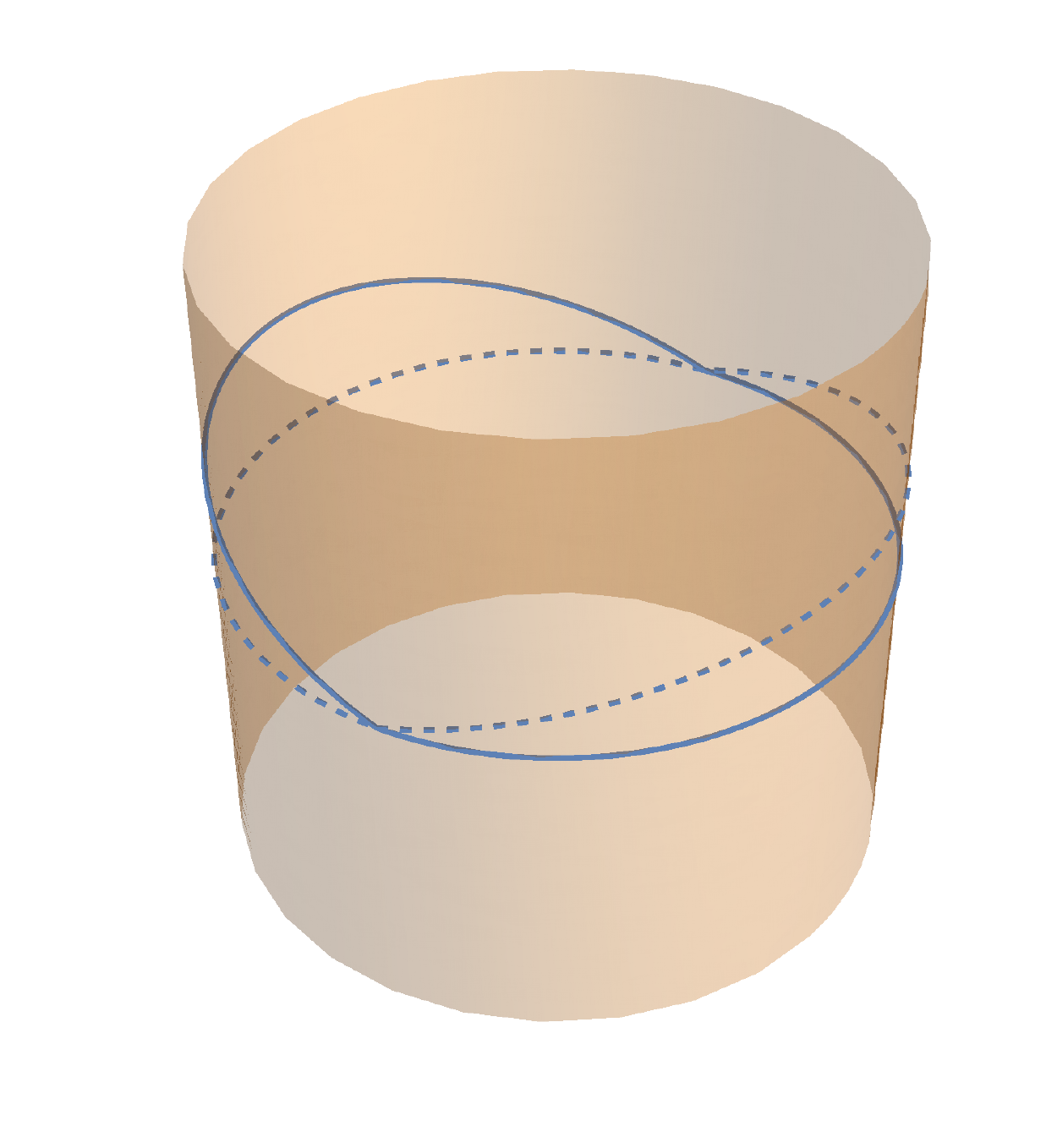}
     \includegraphics[width=4cm]{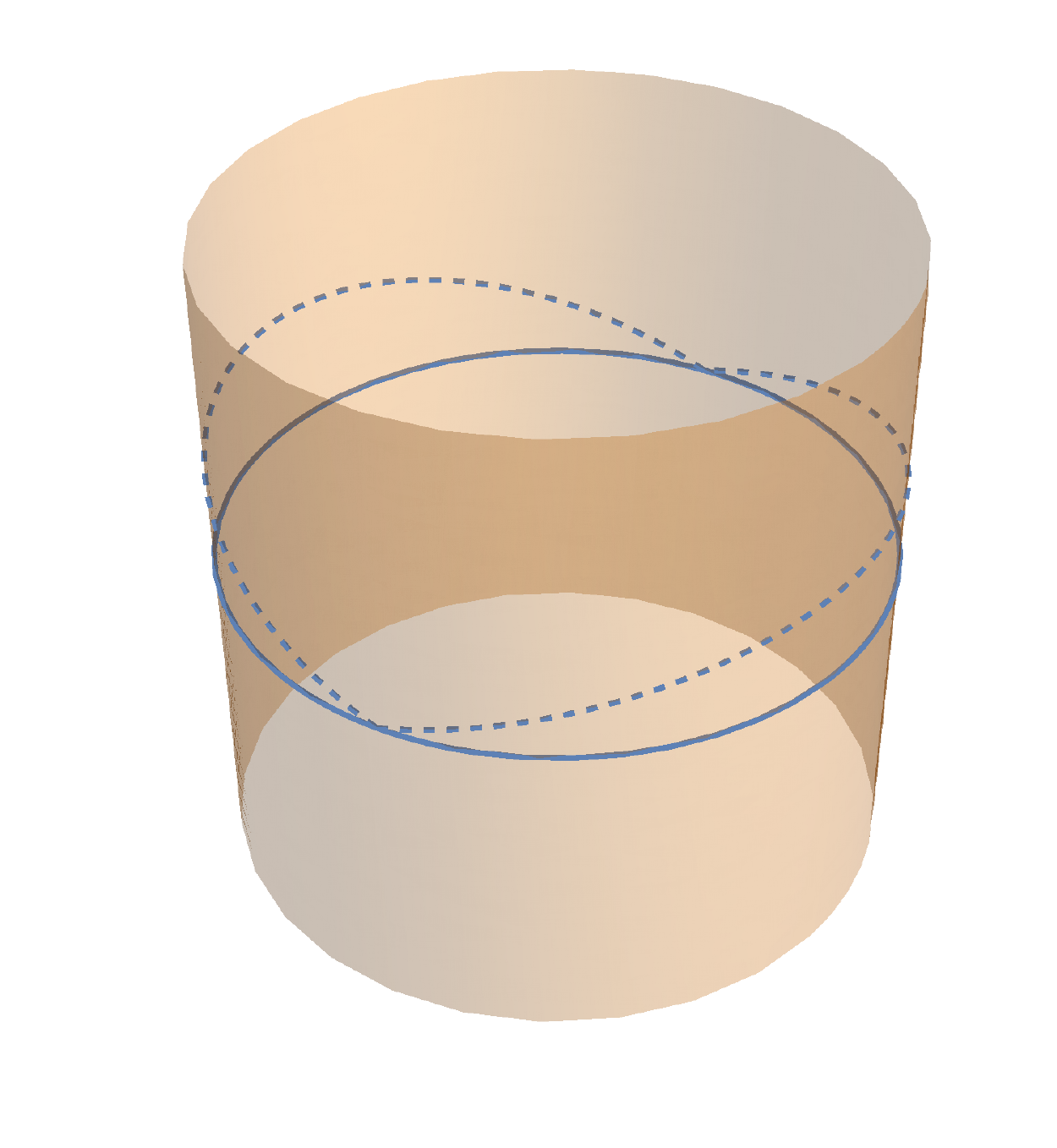}
    \caption{On the left are two Cauchy slices, $\sigma_1$ (in continuous blue) and $\sigma_2$ (in dashed blue), on the boundary of the Poincar\'e patch of an asymptotically AdS spacetime. On the right the corresponding $\sigma_+$ and $\sigma_{-}$ are in dashed blue and continuous blue, respectively.}
    \label{fig:SSA_cauchy_slices}
\end{figure}

This example of a global volume inequality can be generalized to a general bulk spacetime, with one or more boundary components. We illustrate this in Figure \ref{fig:SSA_cauchy_slices}
for a spacetime with one boundary.
Let  $\sigma_1$ and $\sigma_2$ be two Cauchy 
slices of the boundary, and let $\Sigma_1$ and $\Sigma_2$ be the corresponding maximal slices. (As before, we regulate the volume by placing a cutoff surface in the asymptotic region.) Assuming the bulk is time orientable, it admits a foliation by timelike curves, which also extends to the boundary. Each of these curves intersects each of the Cauchy 
slices once. On the boundary define two new piecewise smooth Cauchy 
slices $\sigma_-$ and $\sigma_+$, consisting of the first and second intersection points respectively, and similarly define two new bulk 
slices (which are also only piecewise smooth), $\Sigma_-$ and $\Sigma_+$. Then the boundary of $\Sigma_\pm$ is $\sigma_\pm$,  and 
$\Sigma_\pm$ is generally not the maximal volume slice with this boundary. 
Generalizing the previous notation, let 
 $\mathrm{Vol}(\s)$ denote the maximal volume for a slice bounded by $\s$, and now let $\mathrm{Vol}(\S)$ be the 
 volume of the bulk slice $\S$. Then we have
  $\mathrm{Vol}(\s_\pm)\ge \mathrm{Vol}(\S_\pm)$, and addition of these inequalities yields
$\mathrm{Vol}(\s_+)+\mathrm{Vol}(\s_-)\ge \mathrm{Vol}(\S_+)+\mathrm{Vol}(\S_-)$.
Moreover, $\mathrm{Vol}(\S_+)+\mathrm{Vol}(\S_-)= \mathrm{Vol}(\S_1)+\mathrm{Vol}(\S_2)$, simply because
$\S_+\cup\S_-=\S_1\cup\S_2$, so it follows that 
\beq\label{SSA2}
\mathrm{Vol}(\s_+)+\mathrm{Vol}(\s_-)-  \mathrm{Vol}(\s_1)+\mathrm{Vol}(\s_2)\ge0.
\eeq
To recover the previous case from this generalization, take $\s_1$ to be the two-boundary slice consisting of the union 
of the $t_1$ and $t_4$ slices, and take $\s_2$ to be that consisting of the union 
of the $t_2$ and $t_3$ slices.

\subsubsection{Monotonicity on a boost-symmetric background}
\label{mono}
We next explain how inequality (\ref{SSA1}) implies monotonic growth in time of the complexification rate.\footnote{We thank Adam R. Brown for suggesting the following argument.}
Before showing this, let us discuss the physical significance of this monotonicity property. Black holes are expected to excel at scrambling quantum information and, in particular, should complexify the fastest. Thus, one can expect that their late-time rate of complexification (once transient effects have died out) saturates quantum-information bounds, and in particular should be 
greater than the
complexification rates at earlier times. The monotonicity property coming out of CV-duality is in agreement with this general expectation. By contrast, CA-duality was recently discovered to violate this monotonicity property \cite{Carmi:2017jqz, Couch:2017yil}, perhaps putting into question that particular proposal. 

We now take the the infinitesimal limit of inequality (\ref{SSA1}), setting $t_{1} = t_{L}$, $t_{2} = t_{L} + \delta t_{L}$, $t_{3} = t_{R}$, $t_{4} = t_{R} + \delta t_{R}$. To leading order in small quantities, we find:
\begin{equation}
    \mathrm{Vol}{(t_{L},t_{R})} + \mathrm{Vol}{(t_{L}+\delta t_{L},t_{R} + \delta t_{R})} - \mathrm{Vol}{(t_{L},t_{R} + \delta t_{R})} - \mathrm{Vol}{(t_{L}+\delta t_{L}, t_{R})} = \frac{\partial^{2} \mathrm{Vol}}{\partial t_{L}\partial t_{R}} \delta t_{L} \delta t_{R}.
\end{equation}
Therefore the global inequality (\ref{SSA1}) 
implies positivity of the mixed partial derivative:
\begin{equation}\label{mpd}
    \frac{\partial^{2} \mathrm{Vol}}{\partial t_{L} \partial t_{R}} \geq 0.
\end{equation}
In terms of the new variables
$t_{\pm} = t_{L} \pm t_{R}$, this inequality becomes
\begin{equation}
    \frac{\partial^{2} \mathrm{Vol}}{\partial t_{+}^{2}} - \frac{\partial^{2} \mathrm{Vol}}{\partial t_{-}^{2}} \geq 0.
\end{equation}
For an eternal black hole, the boost symmetry implies that the maximal volume cannot be a function of $t_{-}$ \cite{Stanford:2014jda}. 
We thus end up with the simple statement:
\begin{equation}
    \frac{\partial^{2} \mathrm{Vol}}{\partial t_{+}^{2}} \geq 0.
\end{equation}
which implies that the first derivative of $\mathrm{Vol}$ with respect to $t_{+}$ 
(or equivalently, with respect to either $t_{L}$ ot $t_{R}$ with the other one kept fixed) is monotonic. Replacing the volume by the complexity $\mathcal{C}$, this implies the monotonic increase of the 
complexification rate discussed above.

Note that inequality (\ref{SSA1}) does 
not imply monotonicity of the complexification rate for a general bulk spacetime, since we used the boost symmetry of a 2-sided black hole to deduce it. Nevertheless, we can take the infinitesimal version of the inequality \eqref{SSA2} for a generic spacetime, and derive a condition similar to the positivity of the mixed partial derivative \eqref{mpd}. To this end, consider the case where the two boundary Cauchy slices
$\sigma_1$ and $\sigma_2$ coincide except on two small disjoint bumps to the future, one on $\sigma_1$ 
and one on $\sigma_2$. Expanding to leading order in the size of the bumps, we find that 
the ``off-diagonal" part (since the bumps are disjoint) 
of the second functional derivative of the maximal volume $ \mathrm{Vol}(\s)$ with respect to $\s$ variations is nonnegative.

\section{Quenches, rotation and AdS-Rindler: further probes of CV duality}\label{Sec:FurtherProbes}

CV duality has been primarily probed in the setting of the eternal black hole, where interesting time dependence 
is introduced either by examining foliations that are not Killing time slices, or by introducing shockwave perturbations.
In this section we extend the set of examples, by considering multiple quenches, 
where the black hole temperature
changes, spinning black holes, where the angular momentum provides an extra parameter on which the dependence can be checked, and AdS-Rindler spacetime, where the time dependence of the vacuum complexity in the Rindler wedge is seen to be 
equivalent to that in a black hole spacetime.

\subsection{AdS-Vaidya: event horizon vs apparent horizon}\label{Sec:Vaidya}

The time dependence of holographic complexity has been studied for quenched systems, i.e. systems into which a finite energy density is injected, using the AdS-Vaidya solution \cite{Susskind:2014jwa, Moosa:2017yvt,Chapman:2018dem,Chapman:2018lsv}. 
In this subsection, 
we compare
the growth of the volume inside the horizon for an AdS-Vaidya 
spacetime in the thin shell limit, 
using different definitions of the volume cutoff,
and we extend existing studies to the case of two infalling shells. 

In Section \ref{InsideOut} we discussed several reasons supporting the notion that volume inside the 
black hole horizon is perhaps a more robust measure of complexity of the thermal state than is the volume of a global maximal slice with a cutoff at large distances from the black hole. When the black hole forms from collapse, the degeneracy between different definitions of the horizon is lifted, hence we should examine which (if any) is more appropriate for CV duality. In particular, while the absolute event horizon remains a null hypersurface defined teleologically as the boundary of the past of future null infinity, we shall also consider the apparent horizon, defined here as the boundary of the region containing outer trapped surfaces on the leaves of the maximal foliation
An apparent horizon defined this way is an example of a holographic screen \cite{Bousso:2015mqa,Bousso:2015qqa}, 
i.e. a hypersurface foliated by marginally trapped surfaces. 
Recent work \cite{Engelhardt:2017aux,Engelhardt:2018kcs} 
has shown that the area of a leaf of such a foliation is related to a certain coarse grained holographic entropy, 
which lends support to the idea that the volume inside such surfaces might be directly related to complexity \cite{Zhao:2017isy}.

When the spacetime is time dependent, the maximal time from the horizon 
to the final slice (the ``complexity divisor" of the volume) 
in general becomes time dependent, and in that context it 
might well make more sense to measure the time 
from the apparent horizon rather than
from the event horizon. 
We shall make no attempt here to determine 
which precise extension of the concept is more appropriate.
Instead, we will just address the case where the black hole is either 
the BTZ black hole in $D=3$ dimensions, or in higher dimensions is
large enough so  that the time $\sim \ell$ is
the always the relevant one.

\subsubsection{Single quench}

Consider, then, a black hole formed by an infalling shell in AdS. If the event horizon forms ``at the same time'' as the time on the boundary when the shell starts to fall in, i.e. the time at which an external agent injects some energy into the CFT ground state, then the maximal slice volume inside the horizon remains zero until the injection time, and starts growing after that. This would be consistent with the general expectation that the CFT state starts to complexify after the energy injection.
However, the horizon forms before the injection time if the final horizon radius is greater than the AdS length scale, and after the injection time if it is less \cite{Hubeny:2013dea}.

\begin{figure}[h!]
    \centering
     \includegraphics[width=4cm]{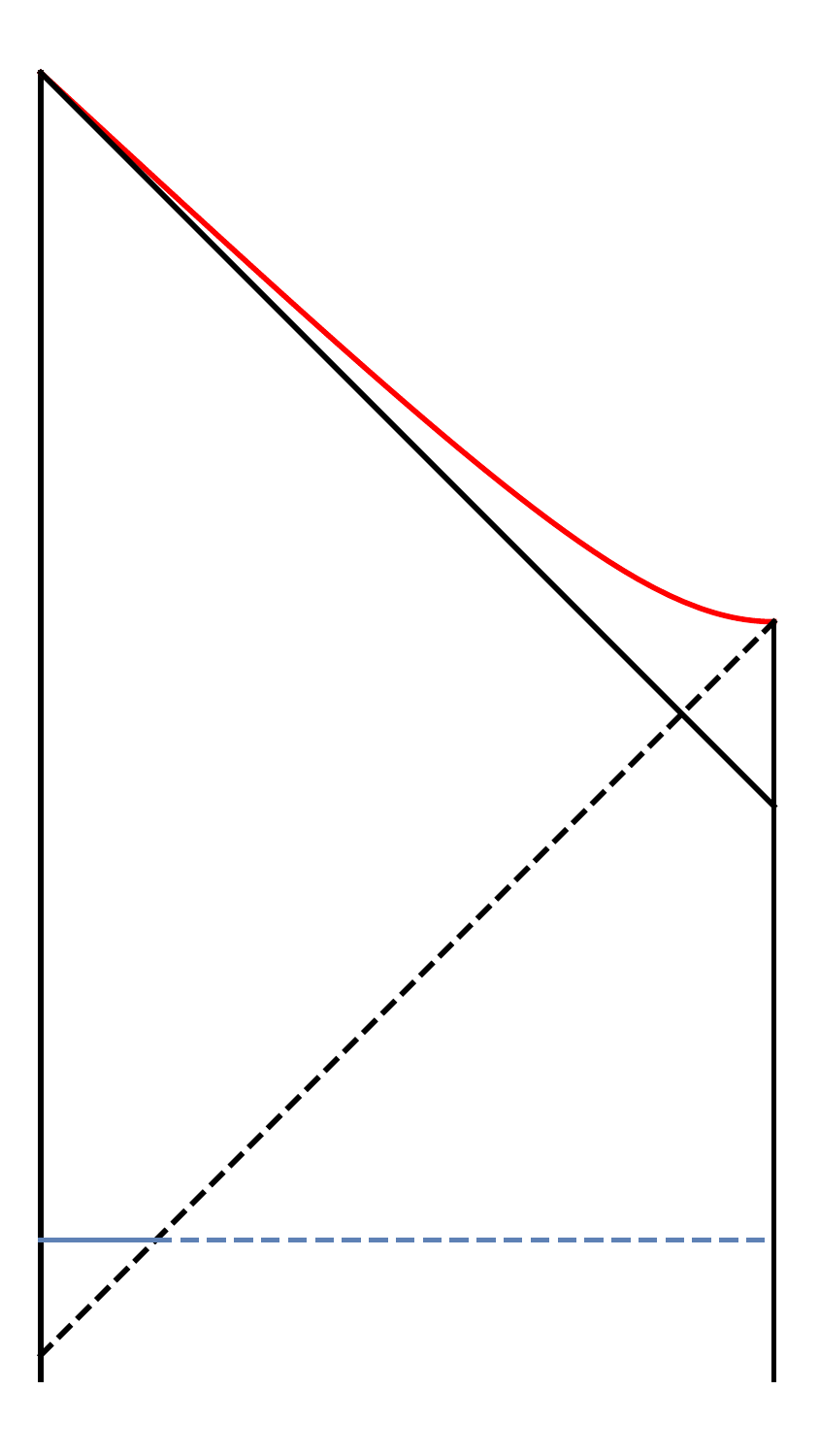}
     \includegraphics[width=4cm]{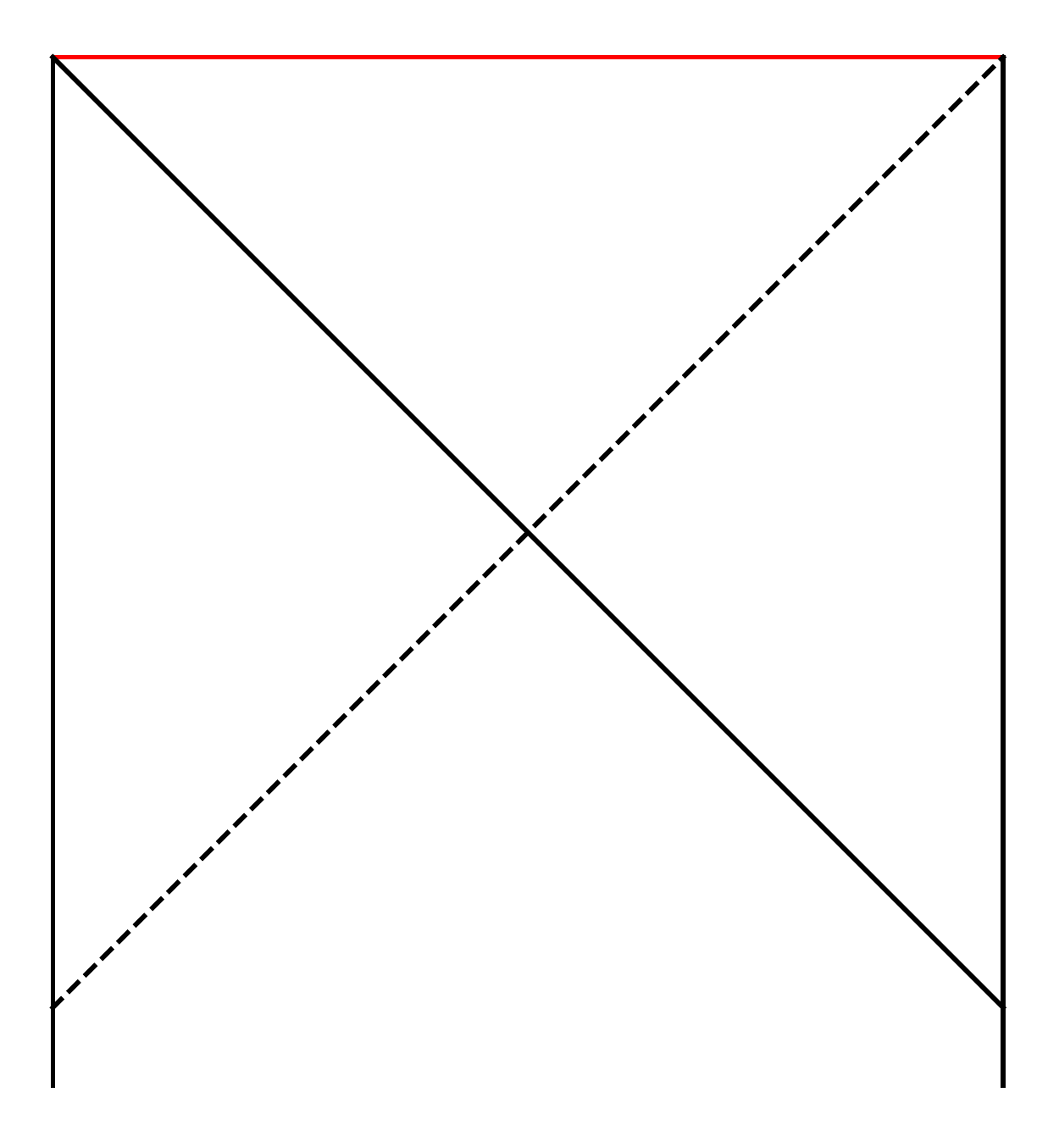}
     \includegraphics[width=4cm]{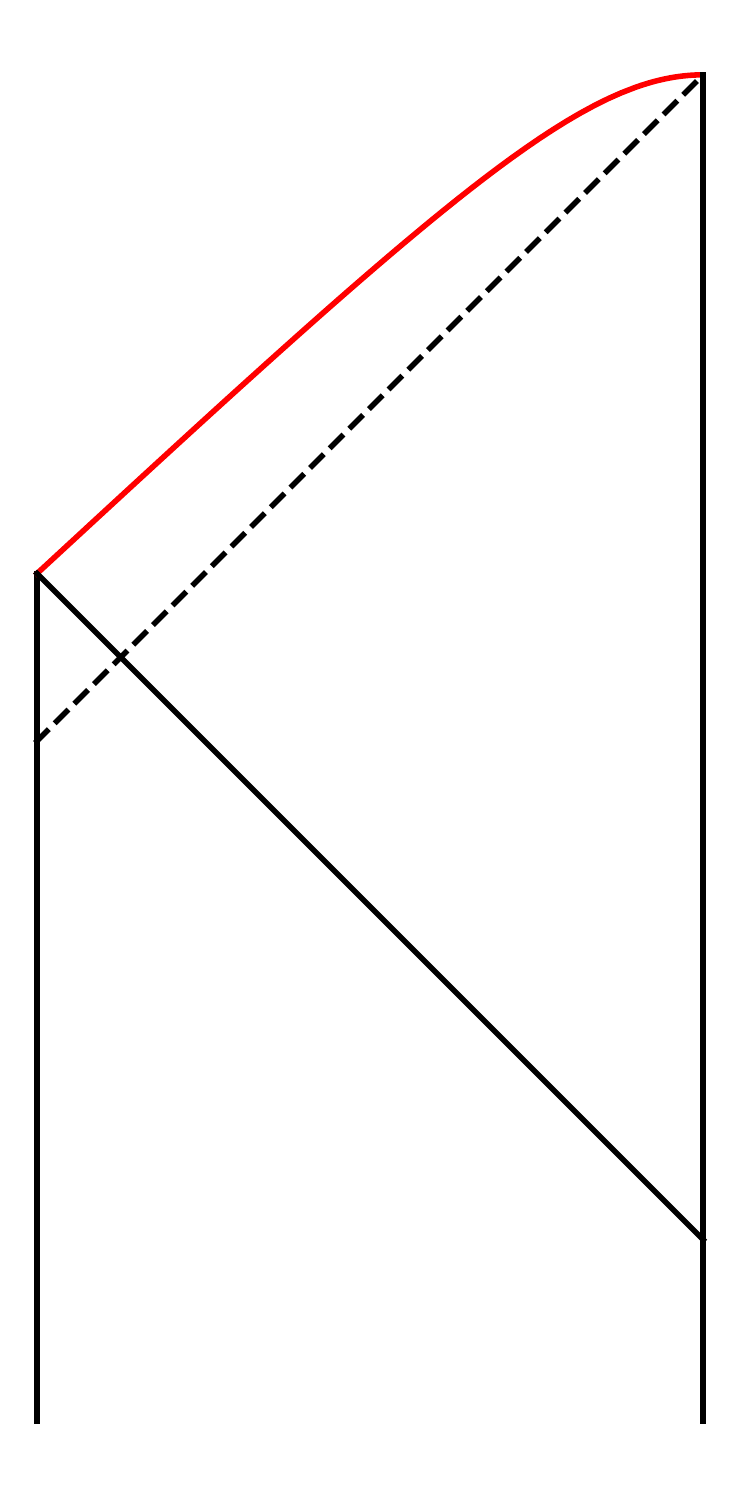}
    \caption{Conformal diagram for the BTZ-Vaidya spacetime in the thin-shell limit for three representative choices of the horizon radius $r_{+}$ (left: $r_{+} = 5L$, center: $r_{+} = L$, right: $r_{+} = L/5$). For all three panels, the center of AdS, the infalling shell and the boundary are in continuous black, the horizon is in dashed black and the singularity is in red. Moreover, on the left panel, we depict a maximal 
    slice anchored at a boundary time to the past of the infalling shell in blue. The portion inside the event horizon is in continuous blue, and the portion outside the horizon is in dashed blue.}
    \label{fig:BTZVaidya}
\end{figure}

To illustrate this, let us work for simplicity in the in the thin-shell limit, and in three spacetime dimensions. 
The BTZ-Vaidya metric in $(r,v)$ coordinates reads:
\begin{equation}\label{BTZVaidya}
    ds^{2} = -f{(r,v)}dv^{2} + 2dvdr + r^{2}d\phi^{2}
\end{equation}
\begin{equation}
    f{(r,v)} = 1 + r^{2} - \Theta{(v)}(1+r_{+}^{2})
\end{equation}
where $\Theta$ is the unit step function. This metric describes a spherical shell at $v=0$ collapsing to form a black hole. To draw the conformal diagram, we need to pass to conformally compactified coordinates $(R,T)$ (see \cite{Hubeny:2013dea} for the coordinate transformation). The metric becomes:
\begin{equation}
    ds^{2} = \frac{-dT^{2}+dR^{2}}{\cos^{2}{R}} + r{(T,R)}^{2} d\phi^{2}
\end{equation}
\begin{equation}
  r(T,R) =
  \begin{cases}
  \frac{(1-r_{+}^{2})\sin{R}-(1+r_{+}^{2})\sin{T}}{2\cos{R}} & \text{if $R+T>0$} \\
  \tan{R} & \text{if $R+T<0$} \\
  \end{cases}
\end{equation}
Fig.~\ref{fig:BTZVaidya} shows the conformal diagrams for three choices of horizon radius (larger than $L$, equal to $L$ and smaller than $L$). The center of AdS is at $R=0$, the boundary is at $R = \frac{\pi}{2}$, and the singularity is at $(1-r_{+}^{2})\sin{R} = (1+r_{+}^{2})\sin{T}$. As illustrated in Fig.~\ref{fig:BTZVaidya}, the horizon forms at the same time $T$ as the injection time on the boundary only in the special case when the horizon radius of the final black hole is exactly equal to the AdS length.

Now consider the large black hole case, and consider a constant $T$ 
slice 
to the past of the infalling shell, where the geometry is locally $AdS$. This 
slice is 
invariant under the $T$ reflection isometry, so it has vanishing extrinsic curvature, hence
is maximal. We depict such a maximal 
slice in blue on the left panel of Fig. \ref{fig:BTZVaidya}. 
Clearly the portion inside the horizon  starts growing even before the energy injection occurs on the boundary.

\begin{figure}[h!]
    \centering
     \includegraphics[width=4cm]{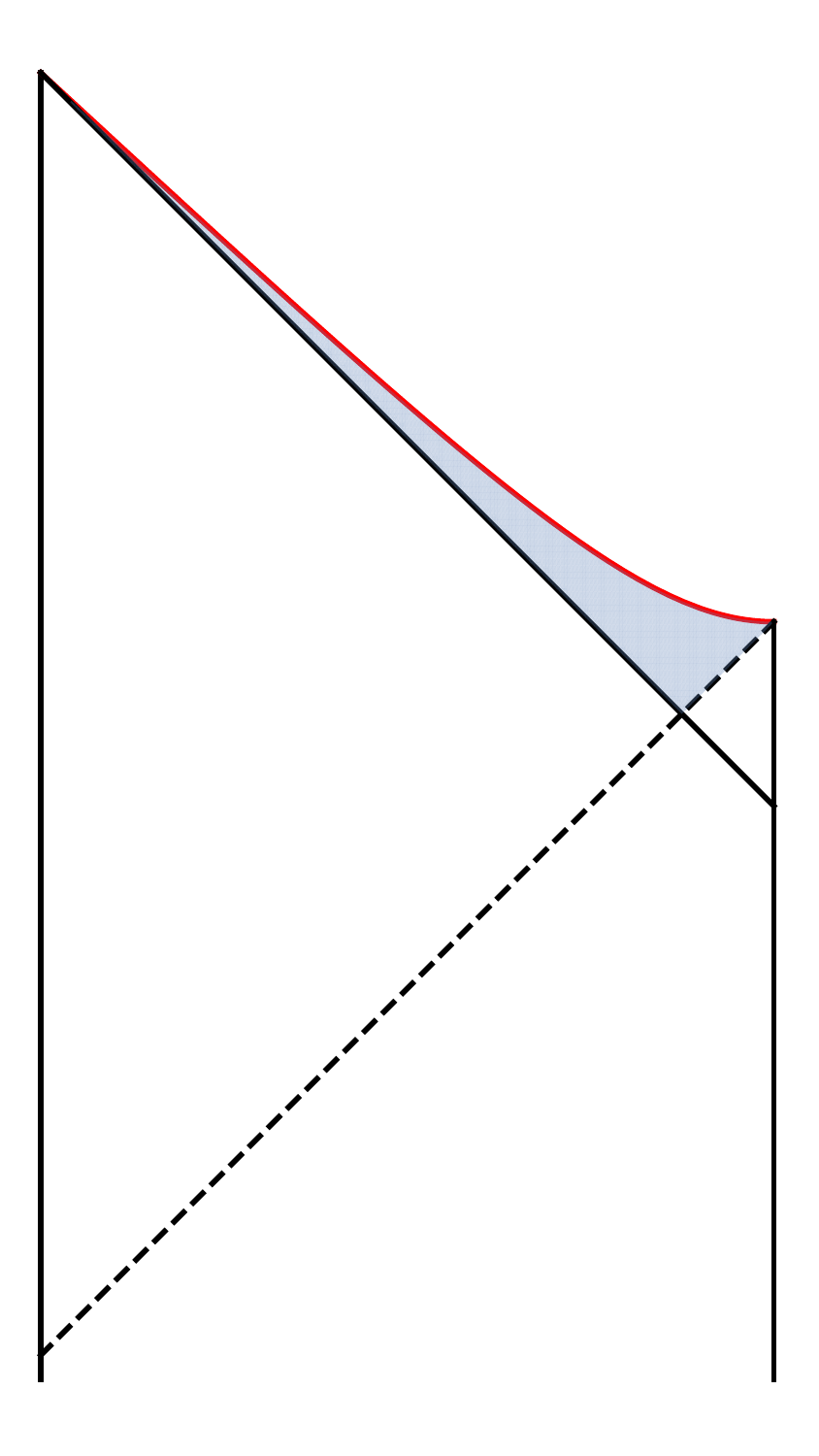}
    \caption{The region inside the apparent horizon is shaded in blue.}
    \label{fig:BTZVaidyaAppHorizon}
\end{figure}
The apparent horizon for Vaidya-BTZ consists of two segments in the conformal diagram (illustrated in \ref{fig:BTZVaidyaAppHorizon}). 
One segment is the event horizon $r=r_{+}$ in the BTZ portion of the spacetime, and the other segment is the infalling shell itself. The interior of the apparent horizon (i.e. the trapped region) is shaded in light blue in Fig. \ref{fig:BTZVaidyaAppHorizon}. 
It is clear that the volume inside the apparent horizon can grow only after the injection occurs, with a delay because the maximal 
slice does not have any portion inside the light blue region of Figure \ref{fig:BTZVaidyaAppHorizon} immediately after injection.

\begin{figure}[h!]
    \centering
     \includegraphics[width=7cm]{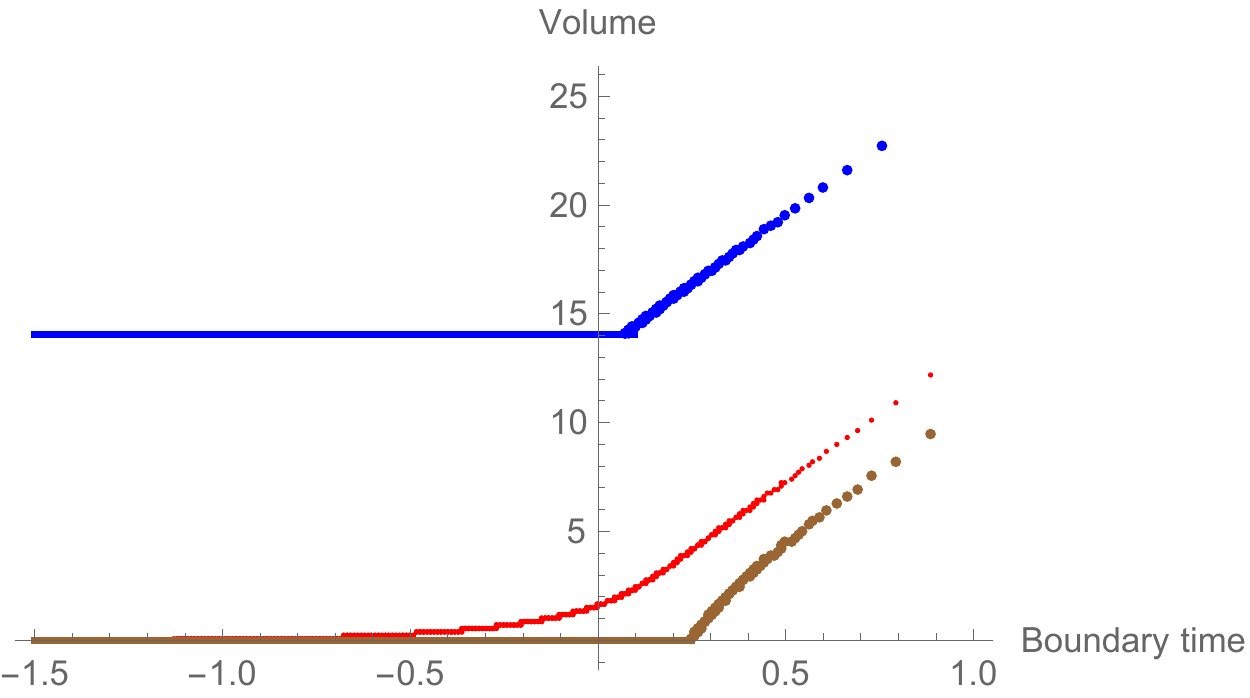}
      \includegraphics[width=7cm]{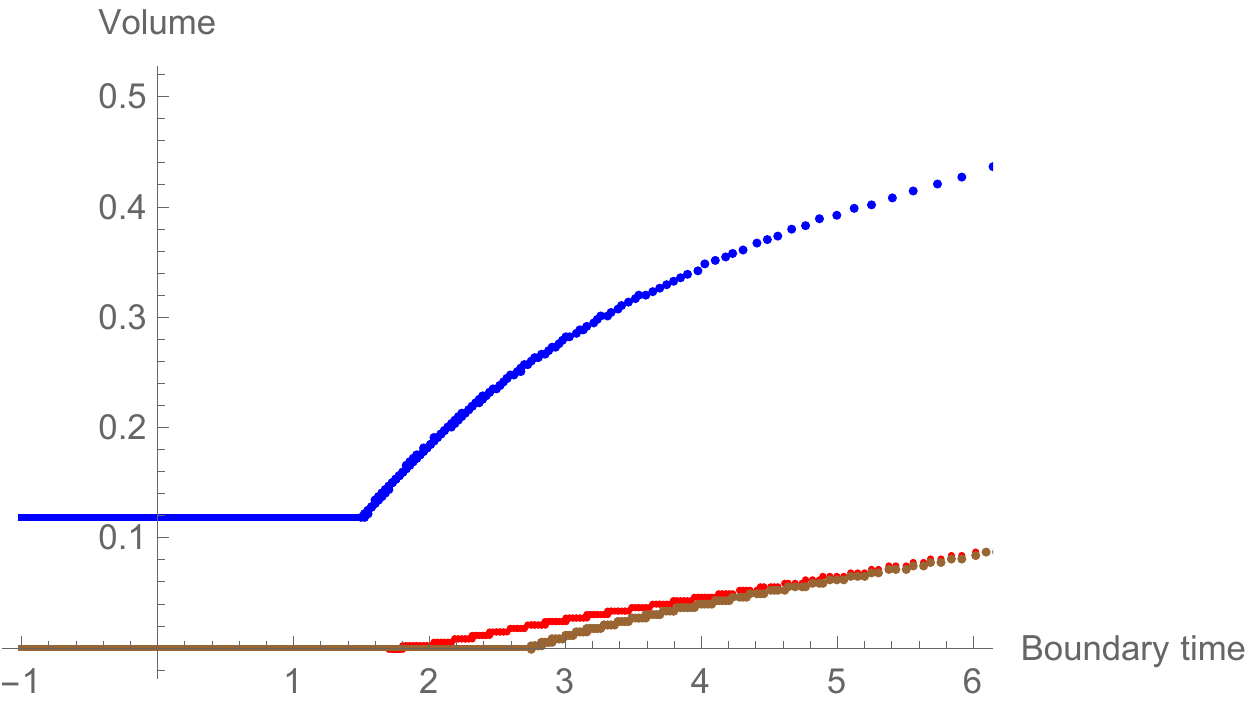}
    \caption{Plot of volume versus boundary time, for a large black hole (left) with $r_{+} = 5\ell$ and a small black hole (right) with $r_{+} = \ell/5$. In both panels, the blue curve is the volume inside some large cutoff, the red curve is the volume inside the event horizon
        and the brown curve is the volume inside the apparent horizon. The boundary time is measured in units of $\ell$.}
    \label{fig:BTZVaidyaVolume}
\end{figure}

In Fig. \ref{fig:BTZVaidyaVolume} we plot the volume inside the horizon as a function of the boundary time,
and compare with the volume inside some large cutoff near the boundary, as well as with the volume
inside the apparent horizon (see below). The plots on the left are for a large black hole with $r_+=5\ell$, while those
on the right are for a small black holes with $r_+=\ell/5$.
(To produce this plot, we solved numerically for the maximal 
slices; the details are explained in Appendix \ref{App:Vaidya}.) 
Several features of this figure are worth remarking upon: 
\begin{itemize}

\item The volume inside a near-boundary cutoff is shown as the blue curve. It starts to grow at the injection time. 
For a large black hole, the volume inside horizon, shown by the red curve, starts growing before the injection time.

\item For a large black hole, the volume inside a near-boundary cutoff (blue curve) grows at essentially the late-time rate 
as soon as the energy is injected. For a small black hole, the growth rate starts out higher, then decreases to to 
late time rate. 

\item The growth rates all converge at late time both for large and small black holes. 
This verifies the expectation that all late time growth of volume occurs inside the horizon 
for a one-sided black hole, as it does for a two-sided one.
    
 \item From a geometrical viewpoint, there are two distinct regimes for the volume inside the event horizon: it may be contained entirely inside the pure AdS part of the spacetime, or it may include a part in the AdS portion and a part in the BTZ portion. 
 There is a critical boundary time at which the former regime transitions to the latter one, on the maximal slice,
 A priori it seems that the volume curve might have a kink at this transition, 
however inspection of the red curves suggests that the derivative is actually continuous at the transition. 
    
 \item The brown curves show the volume inside the apparent horizon, which is always the last to start
 growing. In particular, no growth occurs before injection, and in fact there is a delay between injection 
 and the onset of growth of the volume inside the apparent horizon. 
This delay is perhaps related to a thermalization timescale. As shown in \cite{Balasubramanian:2010ce,Balasubramanian:2011ur}
thermalization takes a time on the order of the AdS length scale, with specific 
behavior depending on the scale at which thermalization 
is probed.

\item The rate of change of the volume inside the apparent horizon starts out higher than the late-time rate, and then approaches the late-time value from above. This is contrary to the monotonic increase expectation. This decreasing behavior was also observed in \cite{Chapman:2018dem}.
Moreover, the longest time to the final slice is the same from all points 
on the outer portion of the apparent horizon (which coincides with the event horizon), and is longer than any other time from inside the horizon, so there is presumably no additional time dependence coming from the 
divisor $\t_f$ in the complexity formula.

\end{itemize}

\subsubsection{Two quenches}

In this subsection
we consider BTZ-Vaidya with two infalling shells. The field theory picture is that energy is injected
twice into the CFT. After the first injection thermalizes, we expect a linear growth of complexity at a rate proportional to the energy injected, since $TS\propto E$. 
After the second injection, the system now has more energy, so we expect the complexification rate to increase. 

If the second injection occurs sufficiently far to the future of the first one, so that the complexity has enough time to reach the linear growth regime before the second injection, CV duality would imply that the
plot of maximal volume versus time will consist of three linear regimes (zero slope, a finite slope, and a bigger finite slope).

The metric is still given by eq. (\ref{BTZVaidya}), except that the function $\Theta{(v)}$ is no longer a unit step function, but a ``double step function'':
\begin{equation}
    \Theta{(v)} = a \theta{(v)} + (1-a) \theta{(v-b)}
\end{equation}
where $a$ is a real number between $0$ and $1$,  $b$ is a positive real number, and $\theta$ is the unit step function. The function $f{(r,v)}$ becomes:
\begin{equation}\label{dubbelstep}
f(r,v) = \left\{
  \begin{array}{ll}
    1 + r^{2} ~~~~~~~&  v < 0\\
    r^{2} - R^{2} & 0 < v < b\\
    r^{2} - r_{+}^{2} &  b < v
  \end{array}
\right.
\end{equation}
with $R^{2} \equiv a(1+r_{+}^{2})-1$. Thus, we have $AdS$ for time $v < 0$, and the usual BTZ with horizon radius $r_{+}$ for $b < v$. In the intermediate regime, $0 < v < b$, we have two qualitatively different cases depending on the sign of $R^{2}$ (it need not be positive): 
If $R^2>0$ we have a BTZ black hole with horizon radius $R$, while if $-1<R^2<0$ we have a ``conical defect'' geometry.\footnote{This conical defect geometry is also a possibility in the single-shell case. This can be seen as follows: the stress-energy tensor of the single-shell AdS-Vaidya in three dimensions is $T_{vv} = \frac{1+r_{+}^{2}}{2r} \Theta'{(v)}$. If $-1 < r_{+}^{2} < 0$, we do not have a black hole final state, yet the null energy condition is still obeyed. This is the conical defect regime.}

\begin{figure}[h!]
    \centering
     \includegraphics[width=4cm]{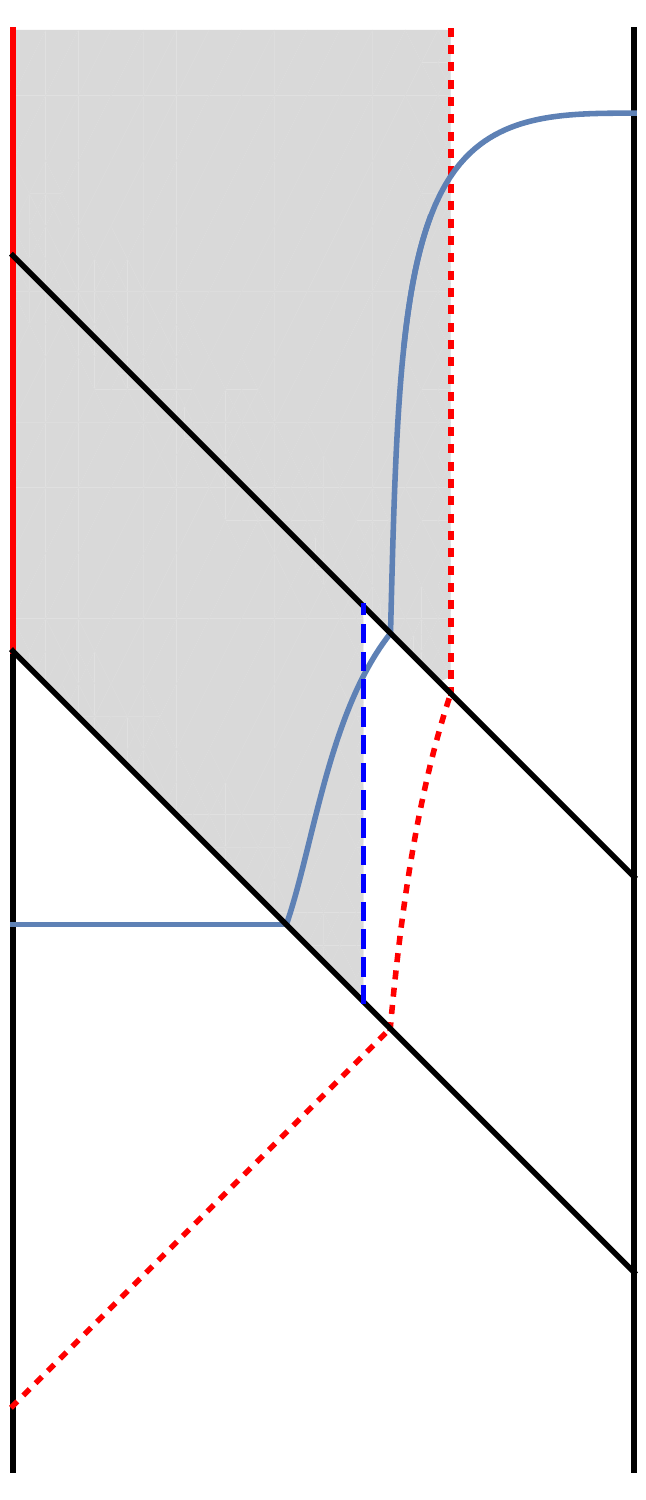}
    \caption{Eddington-Finkelstein diagram of the BTZ-Vaidya solution with two infalling shells. 
    The abscissa and ordinate for the plot are $\rho= \arctan(r)$ and  $t = v -\r$, respectively.
    The two infalling shells, the center of AdS, and the boundary are in thick black, the singularity is in solid red, 
the constant radius portion of the apparent horizon between the shells is in
    dashed blue,  the event horizon 
    is in short-dashed red, and  the maximal 
    slice anchored at late boundary time is shown in continuous blue. 
    The apparent horizon is the boundary 
    of the grey shaded region.
    }
    \label{fig:Vaidya2Shell}
\end{figure}

We have solved numerically for the maximal 
slices and their volume with the parameter choice $a = 1/2$, $b=1$ and $r_{+} = 2$ (see appendix \ref{App:Vaidya} for the technical details). 
The shape of a maximal 
slice anchored at late time on the boundary is depicted using an Eddington-Finkelstein diagram in Fig. \ref{fig:Vaidya2Shell}. 
The effect of the outer shell is to push the 
slice further from the singularity, which
can be explained intuitively as follows: the final slice for the BTZ-Vaidya should approach the final slice of the eternal BTZ black hole (with the same total mass) at late time, which is a constant radius 
slice with
$r\propto r_+$. Since the effect of the outer shell is to increase the horizon radius, it also pushes the final slice further from the singularity.

\begin{figure}[h!]
    \centering
     \includegraphics[width=10cm]{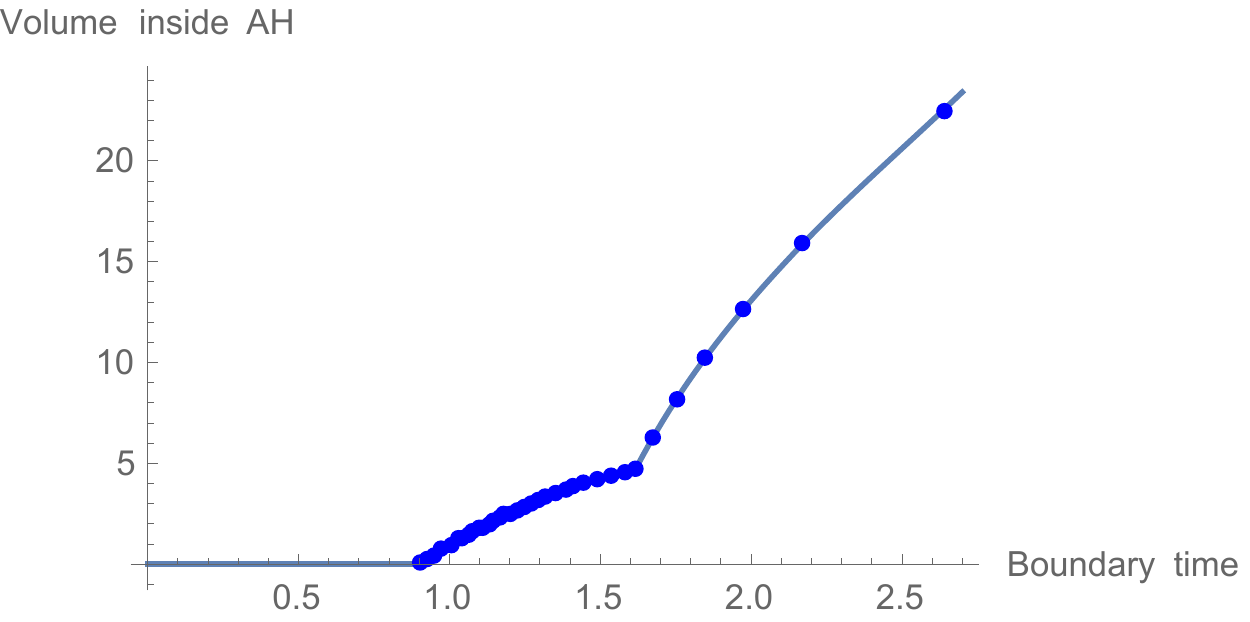}
    \caption{Plot of the maximal 
    slice volume inside the apparent horizon
    in a Vaidya spacetime with two shells. 
   The kinks occur when the maximal 
   slice crosses the points where the first and 
    second shells meet the apparent horizon.}
    \label{fig:Vaidya2ShellVol}
\end{figure}

In Fig. \ref{fig:Vaidya2ShellVol}, we plot the volume of the portion of the maximal slice 
lying inside the apparent horizon
as a function of the boundary time $t_{b}$ at which the 
slice is anchored, with the same parameter choice used for Fig. \ref{fig:Vaidya2Shell}. Note that for a brief period of time after the maximal 
slice crosses the point where
the second shell meets the apparent horizon, the 
slice has two disconnected parts lying inside the apparent
horizon, separated by an annular region falling outside the apparent horizon.

It would be interesting 
to consider similar double quenches but in $D\ge4$ dimensions, with spherical black holes of different sizes, so that the time dependence of the time to final slice divisor $\t_f$ might come into play.

\subsection{Rotating BTZ black hole}
\label{rotating}

A further probe of CV duality is provided by considering rotating black holes.
A rotating black hole is dual to a rotating thermal CFT state. The complexity in such a state should presumably 
grow, with respect to time in the rotating frame, as the entropy of the state times the temperature 
$T_{\rm rot}$ in that frame, since that is the frame in which thermal equilibrium is established. That is,
one would expect that $d{\cal C}/dt_{\rm rot} \sim T_{\rm rot} S$.
While the CFT entropy is frame independent, and is thus equal to the dual black hole entropy, 
$T_{\rm rot}$ is not equal to the dual black hole temperature $T_{\rm BH}$. 
Considering that the thermal frequency defines a clock in the rotating frame, 
there is a time dilation shift of the temperature, and we have $T_{\rm rot} \, dt_{\rm rot} = T_{\rm BH} \, dt_{\rm BH}$, 
where $dt_{\rm BH}$ is the Killing time increment in the asymptotic rest frame of the black hole. It follows that the 
rate of complexity growth can equally be expressed as $d{\cal C}/dt_{\rm BH} \sim T_{\rm BH}S_{\rm BH}$.
We will now show, for the case of a rotating BTZ black hole, that 
CV duality indeed predicts this complexity growth rate at late times.

The metric for the rotating BTZ black hole can be written as
\beq\label{ds2}
ds^2 = -\a^2 dt^2 + \a^{-2}dr^2+ r^2(d\phi -\O_{} dt)^2, 
\eeq
with 
\beq\label{BTZfunctions}
\alpha^2=\frac{(r^2-r^2_+)(r^2-r^2_-)}{r^2\ell^2},\qquad \Omega=\frac{r_-r_+}{r^2\ell},\,
\eeq
and the surface gravity is  $\kappa =(r_+^2-r_-^2)/\ell^2r_+$. The ``final slice," i.e.\ the Killing-invariant maximal 
slice inside the horizon lies where $(r\a)'=0$, which is at $r_f$ given by $r_f^2 = (r_+^2 + r_-^2)/2$. This is a cylindrical surface, 
with induced metric $ds^2 = -\a_f^2 dt^2 + r^2(d\phi -\O_{f} dt)^2$, and volume (area) form $r_f\a_f\, dt\w d\phi$, where 
$\a_f := \sqrt{-\a(r_f)^2}$. The volume of a $dt$ section of the 
slice is thus $dV=2\pi r_f\a_f\, dt$, 
so the rate of change of the total volume inside the horizon, growing at both ends, is $dV/dt = 4\pi r_f\a_f$.
The longest time path from the outer horizon to the final slice has length 
\beq
\t_f=\int_{r_f}^{r_+} \a^{-1} dr = \pi\ell/4,
\eeq
which is the divisor we use in relating the complexity to the volume. 
(Note that, unlike in higher dimensions, this time is independent of the horizon radius, even for small black holes.
In the next section we consider the Kerr black hole in four dimensions, 
and find that one still obtains $TS$ even for small black holes, when the time to the final 
slice is used as the divisor.)

The rate of change of the complexity with respect 
to $t$ is thus 
\beq
\frac{d{\cal C}}{dt} = \frac{4}{\pi\ell} \frac{dV}{dt} = \frac{16r_f\a_f}{\ell}=\frac{8(r_+^2-r_-^2)}{\ell^2}={8\kappa r_+}
=32T_{\rm BH}S_{\rm BH}.
\eeq
As explained above, $T_{\rm BH}S_{\rm BH}$ is equivalent to the rate $T_{\rm rot}S_{\rm rot}$, which 
is what one would expect from a thermal
state \cite{Susskind:2014rva}.
This is a nontrivial check, since the rotating black hole possesses another dimensionless parameter, $r_-/r_+$,
on which the result might have depended. 

We note also that, in the case of the rotating BTZ black hole, the this complexification rate 
from ``complexity = volume'' is interestingly the same as the one found for ``complexity = action'' \cite{Brown:2015lvg}: the late-time rate of growth is proportional to $r_{+}^{2} - r_{-}^{2}$ in both cases.
Here we emphasize the proportionality of this result with $T_{\rm H}S_{\rm BH}$, whereas 
Ref.~\cite{Brown:2015lvg} emphasizes the proportionality with $M - \Omega J$, noting that the late-time rate of growth is slowed down compared to $M$, due to the presence of the conserved charge $J$.

\subsection{Kerr black hole}
\label{Kerr}
Next, we discuss rotating black holes in higher dimensions. This case is substantially more complicated to study than the rotating BTZ one due to the lack of spherical symmetry, and this may be why it has not been studied at all in the literature in the context of CV-duality.\footnote{The volume of constant radius slices inside the Kerr horizon has been studied previously in \cite{Bengtsson:2015zda}.} 
We will consider the case of the Kerr solution in four spacetime dimensions, since the asymptotically flat case is somewhat simpler, and since its maximal 
slices have already been studied to some extent in \cite{Duncan}. 
Our aim is to check whether the complexification rate 
(with the time to the final slice divisor taken into account) continues to be of the order of $T_{\rm H}S_{\rm BH}$.

The Kerr metric for a black hole of mass $M$ and angular momentum $aM$ is given in Boyer-Lindquist coordinates by
\begin{align}\label{Kds2}
ds^2 = - \frac{\Sigma \Delta}{B} dt^2 + \frac{B}{\Sigma} \sin^2 \! \theta \left( d\phi - \Omega dt \right)^2 + \frac{\Sigma}{\Delta} dr^2 + \Sigma d\theta^2,
\end{align}
where
\begin{align}
\Sigma=r^2+a^2 \cos^2 \! \theta, \quad \Delta = \  r^2-2Mr+a^2,\quad %\\ 
B=(r^2+a^2)^2-a^2 \Delta \sin^2 \! \theta, \quad \Omega=2aMr/B.
\end{align}
The inner/outer horizons $r_\pm$ are the roots of $\D=(r-r_+)(r-r_-)$, $r_\pm=M\pm\sqrt{M^2-a^2}$. 

A general axially symmetric maximal  
slice is described by some function $r(t,\theta)$. The final slice is 
a late time limit, so is $t$-independent due to $t$-translation symmetry of the background,
and is therefore described by a function $r(\theta)$
which extremizes the volume. 
The volume element on such a 
slice between the inner and outer horizons (where $\Delta<0$) is 
\beq
\sqrt{\Sigma (|\Delta| - r_{,\theta}^{2})} \sin\theta\, dt\wedge d\wedge d\theta \wedge d\phi
\eeq

As argued in \cite{Duncan}, 
$r(\theta)$ must lie between two values, $r_{\rm min}$ and $r_{\rm max}$,
which are the extrema of $\Sigma\Delta$ with respect to $r$ at $\theta=0$ and $\pi/2$, respectively,
and which are very close to each other for all values of the spin parameter $a/M$. 
The final slice therefore comes very close to being 
a 
slice of constant $r$. We will adopt $r_f\equiv r_{\rm max}$ for that approximate constant value of $r$, 
which is given by
\begin{equation}
    r_{f} = \frac{3}{4}M \left[1 + \sqrt{1 - 8a^{2}/9M^{2}} \right].
\end{equation}
This radius is never parametrically different from the horizon radius $r_+$: 
for $a=0$ it is $3r_+/4$, while for extremal spin it is $r_+$.
The volume element on this cylinder is given by

\beq
\e_f=\a_f\,dt \wedge {\cal A}_f,\quad \a_f=\sqrt{\frac{\Sigma_f|\Delta_f|}{B_f}},
\quad
{\cal A}_f= \sqrt{B_f} \sin\theta\,d\theta \wedge d\phi.
\eeq
The volume of a $dt$ section on the final slice is thus 
$dV=dt \int \a_f{\cal A}_f$, where the integral is over a constant $t$ slice of the cylinder.

Next let us evaluate the maximal 
time to fall from the horizon to the final slice, $\t_f$.
Since the coefficient of $dr^2$ in the line element \eqref{Kds2}
is negative, while those of the other three terms are positive, the
longest time is clearly attained with $dt=d\theta=d\phi=0$. Moreover, the
maximum of these is attained at $\theta=0$, so 
\begin{equation}
    \tau_{f} = \int_{r_{f}}^{r_{+}} dr\;\sqrt{\frac{r^2+a^2}{|\Delta|}}
\end{equation}
Our proposal for the $t$ derivative of the holographic complexity due to growth on one
side of the final cylinder is thus 
\beq\label{CdotKerr}
\frac{d{\cal C}}{dt}=\frac{\int \a_f {\cal A}_f}{\t_f}.
\eeq
Since $r_f\sim r_+$, 
the rate \eqref{CdotKerr} will agree with $\sim T_{\rm H}S_{\rm BH}$ provided 
$\a_f\sim \k{\t_f}$, where $\k$ is the surface gravity. If $\a_f$ were the norm of the horizon generating Killing vector
$\partial_t + \Omega_+ \partial_\phi$ as before, then this last relation would again follow as the first order Taylor expansion, as explained in the introduction.
However, in fact, $\a_f$ is the norm of $\partial_t+\Omega_f\partial_\phi$. Nevertheless,
again, because $r_f\sim r_+$, these two vector fields are not so different, and so
it is plausible that indeed $\a_f/{\t_f}\sim \k$.

To test this relation at the extreme, we define the 
parameter $\epsilon = \sqrt{1 - {a^{2}}/{M^{2}}}$, and expand around extremality, $\e=0$. 
Expanding to lowest order in $\e$, using units with $M=1$, we have $r_\pm=1\pm\e$, $r_f=1+\e^2$,
and $\Delta=(r-1)^2-\e^2$, hence $\Delta_f = -\e^2$. Thus, in computing the volume to lowest order,
we may set $r=1$ in all expressions other than $\Delta$. In particular, 
$\a_f{\cal A}_f\rightarrow \e\sqrt{1+\cos^2\theta} \sin\theta\,d\theta \wedge d\phi$.
At this lowest order in $\e$ we therefore have 
\beq
dV/dt= 2\pi[\sqrt{2}+\sinh^{-1}(1)]\e,
\eeq
and 
\beq
\t_f=\sqrt{2}\int_1^{1+\e} \frac{dr}{\sqrt{\epsilon^2-(r-1)^2}}
%=\int_0^1 ds\, \frac{1}{\sqrt{1-s^2}}
=\frac{\pi}{\sqrt{2}},
\eeq
so our proposal for the complexification rate yields 
\beq\label{1}
\frac{d{\cal C}}{dt}=2\sqrt{2}[\sqrt{2}+\sinh^{-1}(1)]\e\approx 6.49\e.
\eeq
On the other hand, the temperature and entropy  of the Kerr black hole are
\begin{equation}
    T_{BH} = \frac{1}{2\pi} \frac{\sqrt{M^{2}-a^{2}}}{r_{+}^{2} + a^{2}},\qquad
    S_{BH} = \pi (r_{+}^{2} + a^{2}),
\end{equation}
so it follows that 
\beq\label{2}
T_{\rm H}S_{\rm BH} = \frac{1}{2} \sqrt{M^{2}-a^{2}}\rightarrow\half\e.
 \eeq
 
\begin{figure}[h!]
    \centering
\includegraphics[width=8cm]{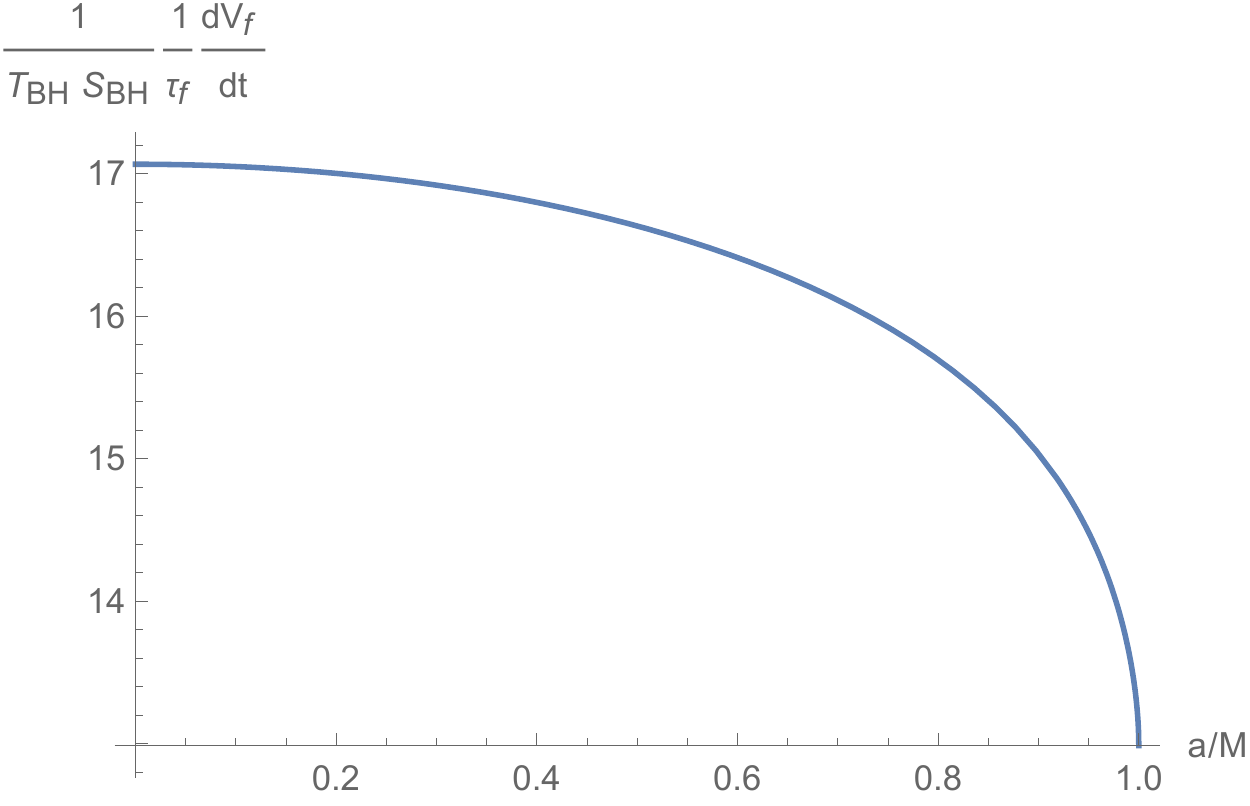}
\caption{Plot of (rate of change of final slice volume $\div$ longest time from the horizon to the final slice)
:(${T_{\rm H}S_{\rm BH}}$), versus the spin parameter for a Kerr black hole in four spacetime dimensions. The ratio is roughly constant over the entire range from
nonspinning to maximal spin.}
\label{fig:KerrFinalSlicePlot}
\end{figure}

The important thing is that \eqref{1} and \eqref{2} are both $O(\e)$, so that their ratio approaches a nonzero pure 
number in the extremal limit. While the ratio depends on the spin parameter $a/M$, it is does not go to zero or infinity
as this parameter goes to zero. 
Finally to exhibit the ratio over the full parameter range, we evaluated it numerically.
As expected, the plot in  
Fig.~\ref{fig:KerrFinalSlicePlot} of the ratio of these quantities
does not vary substantially over the whole range of $a/M$. 
It would be interesting to generalize this analysis to Kerr-AdS, in any spacetime dimension.

\subsection{Rindler wedge complexity growth}
\label{Rindler}

It has previously been observed that many aspects of black hole thermodynamics and horizon entanglement apply to 
acceleration horizons, and specifically to Rindler horizons in flat or AdS spacetimes. In particular, in the AdS/CFT setting, the 
CFT can be partitioned in to two equal halves, and in the ground state the corresponding bulk entanglement wedges are Rindler wedges, separated by a horizon and future and past ``interior" regions analogous to the two-sided black hole interior. Each half of the CFT vacuum is a thermal state with respect to the Hamiltonian generating the conformal boost symmetry of its diamond-shaped domain of dependence \cite{Czech:2012be,Parikh:2012kg}. 
The analogy with the two-sided black hole is close enough that we may expect 
the complexity of the thermal state to grow in time when boosting toward the future on both halves of the partition (as opposed to boosting one side to the future and the other to the past). Moreover, the expected growth rate would be $TS$, where $T$ is the conformal boost temperature and $S$ is the (entanglement) entropy. We now demonstrate that this is indeed the case for 
AdS${}_3$/CFT${}_2$. 

\begin{figure}[h!]
    \centering
     \includegraphics[width=4cm]{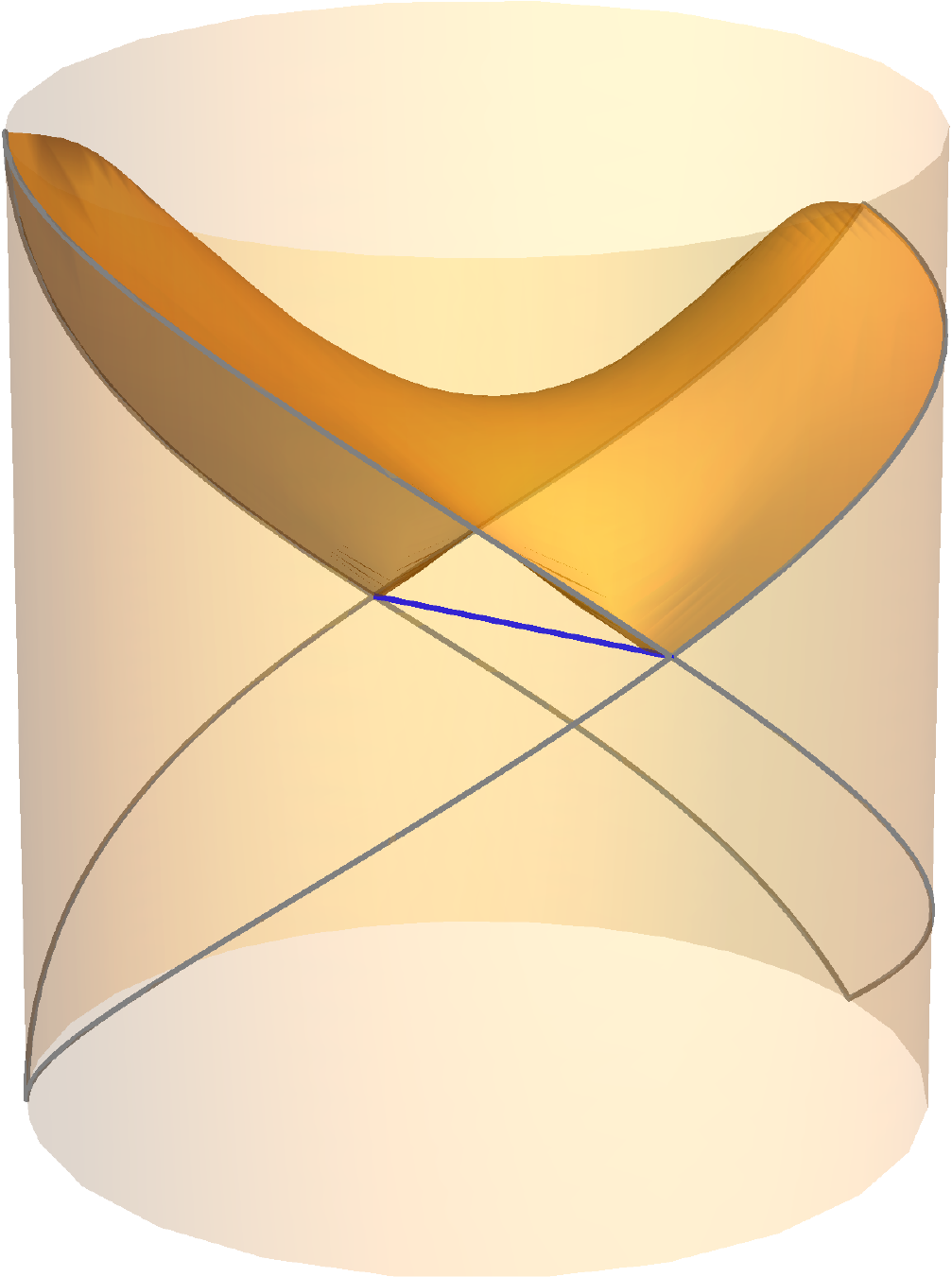}
    \caption{A plot of the final slice of AdS-Rindler embedded into global AdS. Here we use global coordinates for AdS, in which the metric reads $ds^{2} = L^{2}{(-\cosh^{2}{\rho}d\tau^{2} + d\rho^{2} + \sinh^{2}{\rho}d\phi^{2})}$, except that we have compactified the radial coordinate by applying the $arctan$ function (in other words, the boundary of the cylinder is at radius $\pi/2$).
   The line running across the cylinder is the bifurcation line of the Rindler horizon.}
    \label{fig:AdSRindler}
\end{figure}

The metric of AdS in Rindler coordinates is actually just the BTZ metric (\ref{ds2}), with $r_-=0$ and $r_+=\ell$ \cite{Emparan:1999gf}. 
The maximal foliation of interest is defined by constant Rindler time slices of the boundary, and the final slice of this foliation therefore meets the boundary at the Rindler horizon. Fig.~\ref{fig:AdSRindler} displays a plot of this slice in global coordinates,
which lies at $r_f = r_+/\sqrt{2}$ as seen in the previous subsection (\ref{rotating}).
The results of that section also show that the rate of complexification at late times, 
measured with respect to the conformal boost time, is $\propto TS$, where 
$T=1/2\pi$ is the Unruh temperature, and $S$ is the (infinite) Bekenstein-Hawking entropy of the Rindler horizon. 
The dual quantities in the CFT are the conformal boost temperature, and the entanglement entropy of the semicircle, as implied by the Ryu-Takayanagi formula. 
The state on the semicircles is also conformally equivalent to a thermal state on 
 two-dimensional Minkowski space. In higher dimensions it would be conformally equivalent to a thermal state on 
 a static hyperbolic space \cite{Casini:2011kv}.

\section{Conclusion}\label{Sec:Conclusion}
In this paper, we proposed that the apparent lack of universality in the CV duality for large and small black holes
is removed if one identifies the complexity with the volume measured in units of the
maximal (free-fall) time $\t_f$ from the horizon to the final slice
times Planck area.  The distance $c\t_f$ is $\sim$ the AdS radius for spherical black holes large compared to the AdS radius, and it is $\sim$ the horizon radius for small black holes, thus accounting in both cases for the divisor that had been previously 
introduced by hand in order for the complexification rate to match the temperature-times-entropy expectation. We also checked that this prescription matches $TS$ for the rotating BTZ black hole, and the Kerr black hole in four dimensions, for all spin parameters.
While this does seem an improvement over the previous ad hoc assignment, it should be admitted that
we have no reason from first principles for thinking the time $\t_f$ should be relevant, other than that it can be related to the surface gravity and redshift factor at the final 
slice, 
as explained in the Introduction and in Sec. \ref{reason}. Moreover, $\t_f$ is of course 
only defined when a horizon and final slice are present, so is of no use otherwise. 
In this respect, CA duality appears
much more universal. However, it is not so clear whether the notion of complexity and its growth should be expected to have a universal meaning, outside of thermal states, because then the dependence on the arbitrary choice of reference state and gates
with which to define the complexity may be more severe.

We proposed that to capture complexity at the thermal scale one should 
count only the volume inside the horizon, and 
introduced the ``volume current," orthogonal to a foliation of spacetime by maximal 
slices.
This current is a divergence-free vector field, whose flux through the 
slices of the foliation measures their volume. 
This flux picture suggests that there is a transfer of the complexity from the UV to the IR in holographic CFTs, 
which is reminiscent of thermalization behavior decuced using holography. It also 
naturally gives a second law for the complexity when applied at a black hole horizon. 
We further showed how the volume current is a useful tool for establishing
various properties of the volumes of a maximal foliation, 
established a global inequality on maximal volumes that can be used to deduce the monotonicity of the
complexification rate on a boost-invariant background, and probed CV duality in the settings of 
multiple quenches, spinning black holes, and Rindler-AdS.
Finally, we established the existence of a
maximal foliation without gaps (on which the existence of the volume current depends) 
provided that there exists a maximal 
slice anchored at each boundary slice, and
assuming a causality condition, the strong energy condition, and the Einstein equation.

Taken together, these results demonstrate the mathematical and physical utility of the notion of volume 
current associated to a maximal foliation. In the setting of CV duality it is tempting to think of the current as 
a ``gate current" \cite{Headrick:2017ucz}. Perhaps this could be given a more concrete meaning in the context of tensor network models of bulk spacetime.

\section{Acknowledgment}\label{Sec:Acknowledgment}
This work is partially supported by NSF grants PHY-1407744,  PHY-1708139 and PHY-1620610. We would like to thank Vijay Balasubramanian, Adam Brown, Netta Engelhardt, Matthew Headrick, Veronika Hubeny, Brian Swingle and Aron Wall for helpful discussions. 

\appendix

\section{Boundary foliation induces maximal bulk foliation}\label{foliation}

We advertised in Section \ref{Sec:VolumeFlow} that a foliation of the boundary of AdS induces a foliation by globally maximal volume 
slices in the bulk (assuming there exists such a slice terminating on each boundary slice). 
To establish this, we first 
show that, if two boundary slices do not intersect, then the corresponding bulk maximal 
slices\footnote{In this appendix, ``maximal" will always by default mean ``globally maximal".}
do not intersect. 
Next we argue that the (nonintersecting) 
bulk slices fail to be a foliation only if there are two distinct maximal 
slices
anchored at the same boundary slice, and we prove, assuming the strong energy condition
and the Einstein equation, that this cannot happen.
In order to deal with finite volumes, we take the boundary to lie at a finite cutoff surface, which can be taken to infinity at the end. 

\begin{figure}[h!]
    \centering
     \includegraphics[width=6cm]{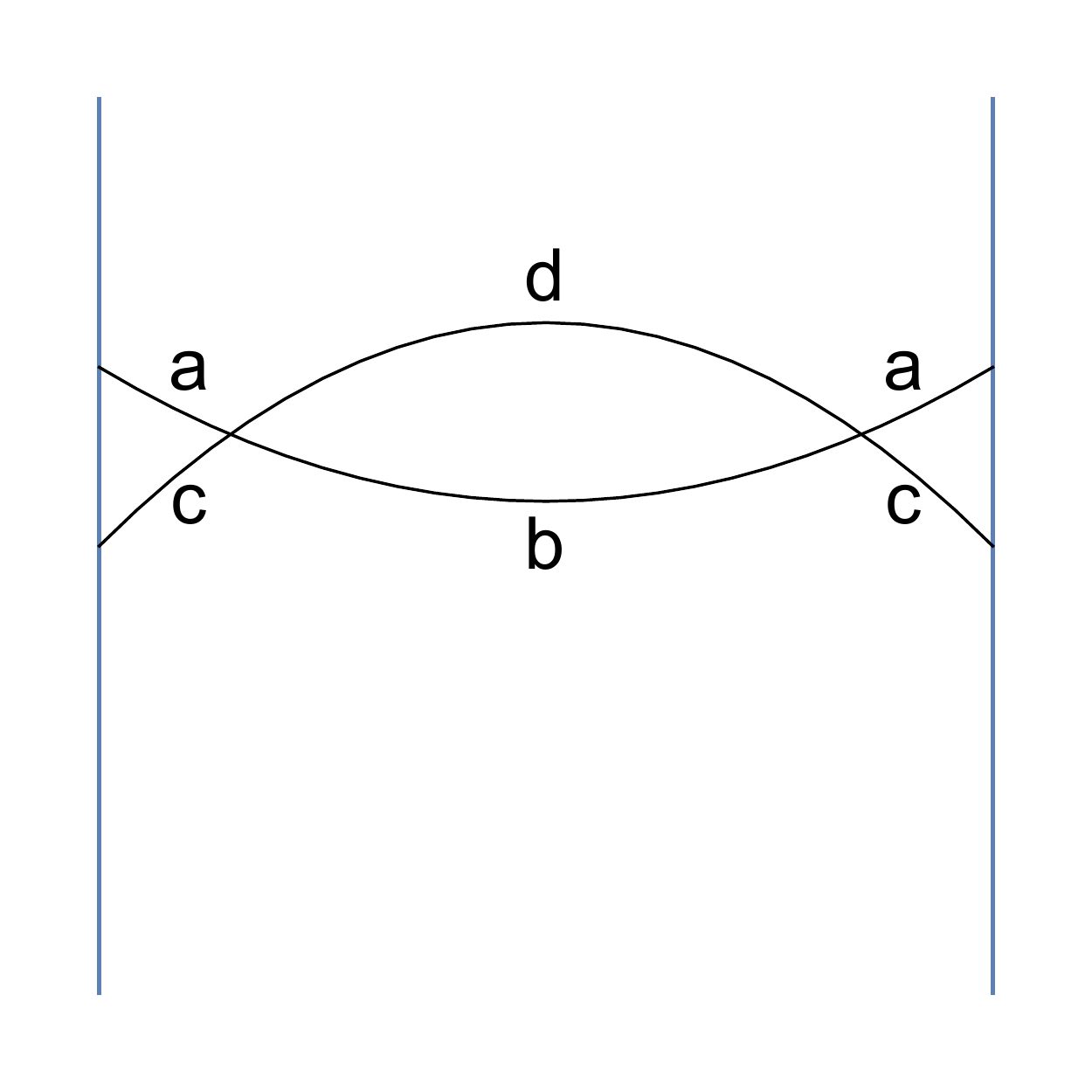}
    \caption{Hypothetical situation where two maximal 
    slices anchored on different boundary Cauchy slice intersect.}
    \label{fig:Contradiction}
\end{figure}

The argument works by contradiction.
Suppose the maximal 
slice anchored at the upper boundary slice dips down sufficiently low in the bulk that it intersects the maximal 
slice anchored at the lower boundary slice (see Figure \ref{fig:Contradiction}). 
Then we can write one maximal 
slice as the union of components $a$ and $b$, and the other maximal 
slice as the union of components $c$ and $d$ as in the figure. 
If $\mathrm{Vol}{(b)} < \mathrm{Vol}{(d)}$, then $\mathrm{Vol}{(ab)} < \mathrm{Vol}{(ad)}$, 
contradicting the maximality of the $(ab)$ 
slice. 
If $\mathrm{Vol}{(d)} < \mathrm{Vol}{(b)}$, then $\mathrm{Vol}{(cd)} < \mathrm{Vol}{(bc)}$, 
contradicting the maximality of the $(cd)$ 
slice. 
The only possibility remaining is $\mathrm{Vol}{(b)} = \mathrm{Vol}{(d)}$.
If that is the case, then $(ad)$ and $(bc)$ would also have to be maximal 
slices. But they cannot be maximal,
since they have corners, and by rounding off the corners their volume can be increased.
If the 
slices are tangent, rather than intersecting transversally,  this ``rounding the corners" argument is not applicable,
but by moving the boundary slices slightly closer together, one would expect that the tangency generically 
becomes a transversal intersection, which would be ruled out by the argument already given.\footnote{This argument 
is essentially an adaptation to Lorentzian signature of a similar argument presented in \cite{Wall:2012uf,Headrick:2013zda} for Euclidean signature in the context of holographic entanglement entropy, establishing the property of ``entanglement wedge nesting" on a static slice.} 
Although not quite a rigorous argument, this seems adequate for our present purposes. 

Now if the bulk maximal 
slices do not intersect, then the boundary foliation will induce a bulk foliation unless there
are gaps where the family of maximal volume 
slices jumps discontinuously across some spacetime region.
Since the metric is assumed continuous, however, the maximal volume function itself cannot jump discontinuously as the boundary 
slice is pushed toward the future. Hence, if a gap does occur there must be two maximal  
slices with the same volume, anchored at the same boundary slice. We now argue that this cannot happen, 
given a causality assumption, the Einstein equation, and the strong energy condition.
In fact, the argument will establish a stronger result: there cannot be two extremal bulk slices with the same boundary. 

Suppose there are two such slices, $\S_1$ and $\S_2$, with $\S_2$ to the future of $\S_1$, with the same, co-dimension-2 boundary, and both with ${\rm Tr}K=0$. While the domains of dependence ${\cal D}_1$ and ${\cal D}_2$ of $\S_1$ and $\S_2$
are each automatically globally hyperbolic, we need to assume that $\S_2\subset {\cal D}_1$ and 
$\S_1\subset {\cal D}_2$, which amounts to assuming that the domains of dependence coincide, ${\cal D}_1={\cal D}_2$. This condition ``obviously" holds for ``normal" causal structures. Under this causality assumption, we can invoke Theorems 9.4.3 and 9.4.5 and Lemma 8.3.8 of Ref.\cite{Wald} to infer that every point $p$ on $\S_2$ lies on a geodesic that maximizes the proper time from $p$ to $\S_1$, meets $\S_1$ orthogonally, and has no conjugate points between $p$ and $\S_1$.  The congruence 
of these geodesics maps (possibly a subset of) $\S_1$ onto all of $\S_2$. The expansion $\theta$ of the congruence at $\S_1$ is equal to ${\rm Tr}K$, which vanishes by assumption. The Raychaudhuri equation together with the timelike convergence condition or, assuming the Einstein equation, the strong energy condition, then implies that $\theta$ is decreasing everywhere along the congruence.\footnote{Strictly speaking, we need here to assume the generic condition, that $R_{ab}u^au^b\ne0$ somewhere along each geodesic, where 
$u^a$ is the geodesic tangent. In the case with a (negative) cosmological constant, this is automatic.}
Moreover,  $\theta$  cannot go through $-\infty$ before reaching $\S_2$ since, 
as stated above, the time-maximizing curve has no conjugate points between $p$ and $\S_1$.
It follows that 
$\theta$ is negative everywhere, which implies that the geodesic flow is volume-decreasing. That is, the volume of a small ball carried along by the flow will {\it decrease}, as measured in the local rest frame of the flow. Furthermore, since the geodesics do not generally meet
$\S_2$ orthogonally, the volume of a small patch of $\S_2$ on which the flow lands will be {\it less} than the volume of the small ball
carried by the flow. It follows that the volume of $\S_2$ is less than the volume of the pre-image of $\S_2$ in $\S_1$ under this flow, and {\it a fortiori} the volume of $\S_2$ is less than that of $\S_1$. Similarly, we can argue the opposite, and thus we reach a contradiction, since the volume of $\S_2$ cannot be both less than and greater than that of $\S_1$. The initial assumption is therefore false: there cannot be two extremal slices with the same boundary. 
Together with the previous results, this implies that a boundary foliation determines a maximal bulk foliation without gaps.
Note that the latter need not completely cover the bulk, however. For example,  as discussed in the text, the maximal 
slices for a two-sided black hole do not extend beyond a final slice, located inside the event horizon.

We established the uniqueness property of extremal 
slices with a given boundary 
using a global argument in which the existence of time maximizing curves without conjugate points
played a key role. However, it was briefly mentioned by Witten, in a conference talk \cite{WittenAdS20}, 
that uniqueness can be proved in a different fashion, namely,
by (i) showing that the volume of any extremal 
slice is a local maximum with respect to small deformations, and (ii) arguing that if there were two local maxima, there would necessarily also be a saddle point of the volume, contradicting the fact that all extremal 
slices are local
maxima of the volume. The reasoning for point (i) is simple and local:
the expansion of the congruence of timelike  geodesics orthogonal to any extremal 
slice starts out zero at the 
slice, and the strong energy condition (together with the Einstein equation) implies that it is negative and decreasing off the 
slice. The transversal spatial volume therefore decreases along the congruence, and non-orthogonality of the congruence to the deformed 
slice implies that the
latter has even smaller volume, so the extremal 
slice  is a local maximum of volume.
The reasoning for (ii), the existence of the saddle point, 
was not as explicit in the talk, but it was pointed out in a picture that there would necessarily be 
a local minimum along some one-parameter family of 
slices joining them. This is of course 
a necessary condition for the existence of a saddle point, but it is not clear to us that 
a saddle point is guaranteed to exist.

We end this appendix with an example where the strong energy condition does
{\it not} hold, and consequently there can be 
more than one extremal hypersurface with the same boundary, and 
the extremal 
slices are {\it not} local maxima of volume: de Sitter spacetime.
The constant time slices of a static patch of de Sitter spacetime 
are all anchored at the same location on the boundary 
of the patch, all are extremal, and all have the same volume. 
To visualize this, consider the two dimensional de Sitter hyperboloid 
embedded in three dimensional Minkowski spacetime. The constant-time slices of the static patch are equatorial semicircles on the de Sitter hyperboloid, 
and are related to one another by Lorentz boosts in the embedding spacetime.

\section{Techniques to evaluate the maximal volume}\label{App:Techniques}

In this appendix, we demonstrate the use of the flux picture of complexification as a technical tool for explicit computation. The techniques presented here complement existing studies in the literature such as \cite{Carmi:2017jqz}, where the maximal volume was computed by maximizing the volume functional directly. In subsection (\ref{App:BTZFlux}), we evaluate the volume flux for the BTZ black hole. In subsection (\ref{App:NullFlux}), we present a variation of this technique when the cutoff is null, in which case the flux density is given by the lapse function. 

\subsection{Direct evaluation of flux}\label{App:BTZFlux}
In this appendix, we present the derivation of the volume current and the volume flux for the BTZ black hole. The boundary foliation is the symmetrical one $t_{L} = t_{R}$.

Let us start with the BTZ black hole. We work in $(r,v)$ coordinates, which are regular across the horizon:
\begin{equation}\label{BTZmetric}
    ds^{2} = -f{(r)}dv^{2} + 2dvdr + r^{2}d\phi^{2}
\end{equation}
\begin{equation}
    f{(r)} = \frac{r^{2}-r_{+}^{2}}{L^{2}}
\end{equation}
The function $v(r)$ describing the shape of the maximal 
slices was essentially worked out in \cite{Carmi:2017jqz}. Its derivative is given by:
\begin{equation}\label{dvdrBTZ}
\frac{dv}{dr} = \frac{\sqrt{f(r)r^{2}+C^{2}}-C}{f(r)\sqrt{f(r)r^{2}+C^{2}}}
\end{equation}
where $C$ is a positive constant.\footnote{$C$ is the negative of the ``energy'' $E$ in \cite{Carmi:2017jqz}. From the viewpoint of that paper, the constant $C$ arises as a ``conserved quantity'' associated with the $v$-independence of the volume functional.} The constant $C$ labels the particular maximal 
slice in the foliation, and it ranges from $0$ (for the 
slice anchored at $t_{L} = t_{R} = 0$) to $r_{+}^{2}/2L$ (for the final slice).\footnote{To see this, note that $\frac{dv}{dr} = \frac{1}{f}$ on the slice $t_{L}=t_{R}=0$ (since this slice is at $t=0$). As for the final slice, symmetry dictates that it is a slice of constant $r$, and $\frac{dv}{dr}$ diverges. Both of these facts are verified by plugging in $C=0$ and $C = \frac{r_{+}^{2}}{2L}$ respectively.} The unit normal 1-form to the slice labeled by $C$ is:
\begin{equation}\label{BTZunitnormal}
    n_{\mu}dx^{\mu} = -\sqrt{f + \frac{C^{2}}{r^{2}}} dt - \frac{C}{rf} dr
\end{equation}
To get the volume current $v^{\mu}$, we glue together the unit normal to all the slices labeled by different values of $C$. This amounts to promoting $C$ to the function of $v$ and $r$ implicitly given by integrating (\ref{dvdrBTZ}) from the midpoint (also called the ``throat'' in the numerical relativity literature) outward:
\begin{equation}\label{vofrBTZ}
    v = -\frac{1}{r_{+}}\mathrm{arctanh}{\left( \frac{r_{C}}{r_{+}} \right)} + \int_{r_{C}}^{r} \frac{\sqrt{f(x)x^{2}+C^{2}}-C}{f(x)\sqrt{f(x)x^{2}+C^{2}}} dx
\end{equation}
The first term on the right-hand side is the tortoise coordinate of the throat, and $r_{C}$ is the radius of the throat given by:
\begin{equation}
    r_{C} = \sqrt{\frac{1}{2} (r_{+}^{2} + \sqrt{r_{+}^{4} - 4C^{2}L^{2}})}
\end{equation}
The two equations above define $C(r,v)$. We then find the volume current:
\begin{equation}
    v^{\mu}\partial_{\mu} = \frac{1}{f(r)} \left( \sqrt{f(r) + \frac{C(r,v)^{2}}{r^{2}}} - \frac{C(r,v)}{r} \right) \partial_{v} - \frac{C(r,v)}{r} \partial_{r}
\end{equation}
It can be checked that both components of $v^{\mu}$ are regular at the horizon. The volume element is $\mathcal{\epsilon} = r dv \wedge dr \wedge d\phi$. Computing the interior product $v \cdot \mathcal{\epsilon}$ and restricting to the cutoff at constant $r = r_{c}$ yields:
\begin{equation}
    v \cdot \mathcal{\epsilon} \bigg|_{r_{c}} = C(r_{c},v) dv \wedge d\phi
\end{equation}
Note that $r_{c}$ is allowed to be the horizon since our formalism can handle null surfaces. Evaluating the flux, we then find the change in the volume between $v_{1}$ and $v_{2}$ to be:
\begin{equation}
    \Delta V = 2\pi \int_{v_{1}}^{v_{2}} C{(r_{c},v)} dv
\end{equation}
In the usual near-boundary cutoff $r_{c} \rightarrow \infty$, the $v$ coordinates becomes the boundary time coordinate $t$ and the function $C(r_{c},v)$ is nothing but the flux density, or the complexification rate. The main lesson from this computation is that the flux density coincides with a certain time function $C$ for the maximal slicing.

We also note that it is possible to work with Schwarzschild coordinates $(r,t)$ instead of $(r,v)$, despite the coordinate singularity at the horizon. In Schwarzschild coordinates, the shape of the maximal 
slice reads:
\begin{equation}
    t = -\int_{r_{C}}^{r} \frac{C}{f{(x)}\sqrt{f{(x)}x^{2}+C^{2}}} dx
\end{equation}
To integrate across the horizon, we should understand the integral above in the sense of the Cauchy principal value \cite{Estabrook}. In fact, the plot (\ref{fig:VolumeFlowDisplay}) of the volume flow was generated by working in Schwarzschild coordinates and using the Cauchy principal value to continue the maximal 
slice across the horizon.

Finally, it might appear surprising that the flux of the volume flow yields a finite answer when the cutoff is taken to the boundary, especially if we think about the flow direction near the boundary. The maximal 
slice should become tangential to the constant Killing time slices near the boundary, and since the volume current is orthogonal to the maximal 
slices, it may seem that the volume flux across a constant-$r$ cutoff is zero as $r \rightarrow \infty$. That this is not the case can be 
understood as follows: at a finite but large cutoff in the bulk, the flow direction has a small component not orthogonal to the constant Killing time slice. As the cutoff is sent to the boundary, this small component tends to zero, but the volume element on the cutoff also diverges in the same time. This divergence cancels the vanishing of the subleading component in a way to yield a finite answer.

\subsection{Flux across a null surface}\label{App:NullFlux}

In this appendix, we elaborate on the particular case of the volume flux across a null surface, and relate this flux to the lapse function {for a time function $\t$ defining the maximal foliation.
Recall that to a time function $\t$ we can associate a \textit{lapse function} $N$ defined by:
\begin{equation}
    n= N d\t
\end{equation}
where $n$ is the unit normal 1-form to a constant $\t$ slice, with sign chosen so that $N>0$.  
Since the volume current vector $v$ is the unit normal vector to the maximal 
slices, 
we have $v\cdot n=1$.
Now let $k$ be the null normal 1-form to the horizon, normalized so that $v\cdot k = 1$, and let $\A$ be the area form of the intersection of the maximal slice with the horizon,
so that $\epsilon = k\w n \wedge \A$. The volume current 
is then $v \cdot \epsilon = n \wedge \A - k \wedge A$, whose 
pullback to the null surface is $n \wedge A = N d\tau \wedge \A$, since the pullback of $k$ vanishes. 
The  2-form $N \A$  thus serves as the volume flux density.}

\begin{figure}[h!]
    \centering
     \includegraphics[width=7cm]{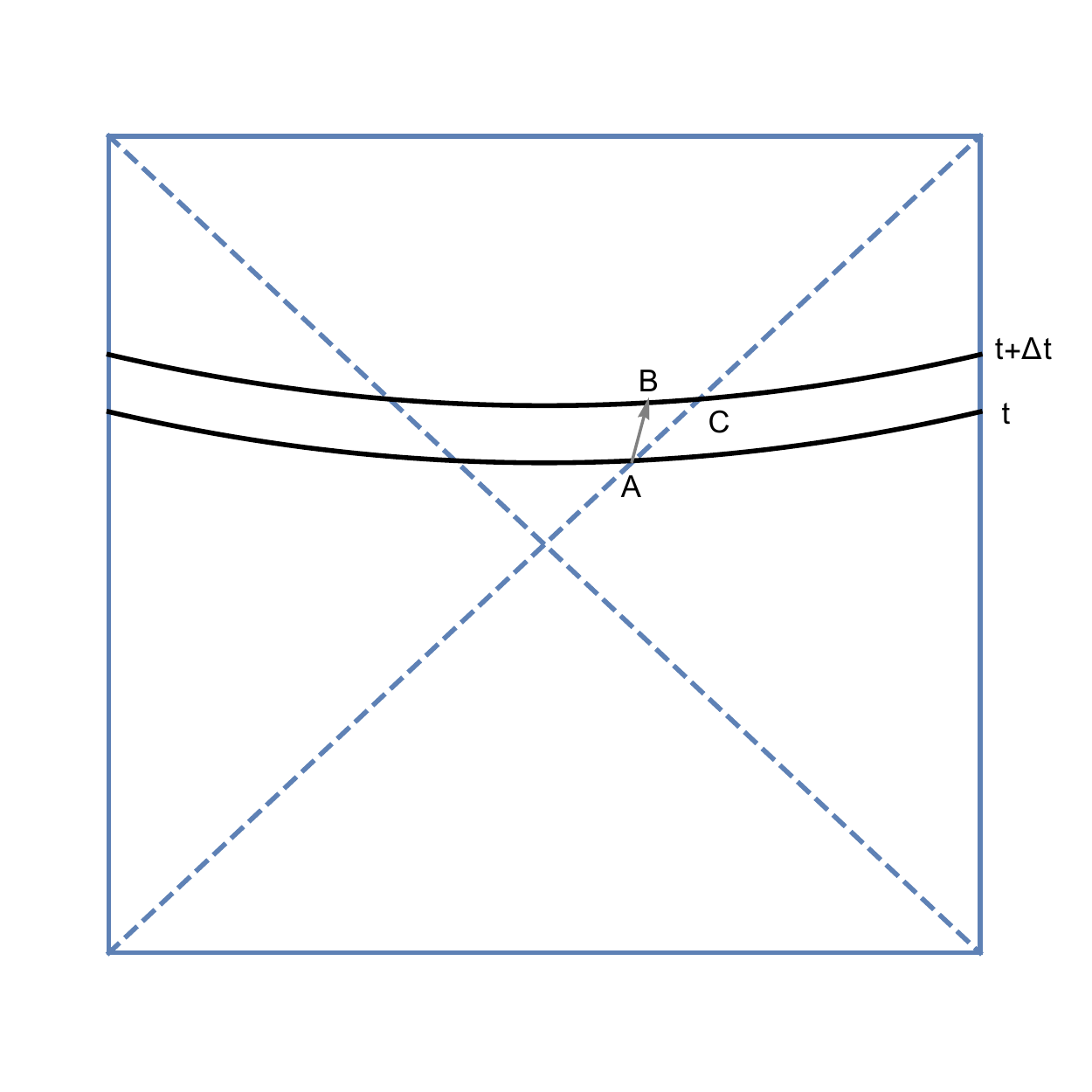}
    \caption{Illustration of the derivation of (\ref{Flux&Lapse}).}
    \label{fig:HorizonFlux}
\end{figure}

This fact can be seen more {geometrically} 
as depicted in Figure (\ref{fig:HorizonFlux}). We pick two slices in the foliation labelled by 
$\t$ and $\t+\Delta \t$, with  $\Delta \t$ small. Let $A$ be the intersection between slice $\t$ and the horizon, and let $C$ be the intersection between slice $\t+\Delta \t$ and the horizon. Moreover, consider {a volume flow worldline} passing through $A$, and let $B$ be the intersection of that worldline with the slice $\t+\Delta \t$.
Since $\Delta \t$ is infinitesimal, the {geometry of the} triangle $ABC$  
{is like} in flat space. {Since $AB$ is orthogonal to $BC$, }
we conclude that $AB = BC$, where $AB$ denotes the proper time elapsed along the worldline between $A$ and $B$, and $BC$ denotes the proper length of the segment of the slice $\t+\Delta \t$ between $B$ and $C$. 

Now consider the increment in the volume $\Delta \mathrm{Vol}$ between $\t$ to $\t+\Delta \t$. We have $\Delta \mathrm{Vol} = 4\pi r_{+}^{2} BC = 4\pi r_{+}^{2} AB$, where we work in 3+1 dimensions for concreteness, and in the second equality we used the relation derived in the previous paragraph. (There is also an identical contribution from the left-side of the Penrose diagram, which we ignored.) On the other hand, we have $AB = N{(A)}\Delta \t$. Thus, we can relate the volume increment to the lapse as follows:
\begin{equation}
    \Delta \mathrm{Vol} = 4\pi r_{+}^{2} N{(A)} \Delta \t.
\end{equation}
For a finite time difference, we integrate the lapse:
\begin{equation}\label{Flux&Lapse}
    \Delta \mathrm{Vol} = 4\pi r_{+}^{2} \int_{\mathcal{H}} N d\t.
\end{equation}
This is the formula we are after: the volume flux across the horizon is also the integral of the lapse along the horizon. In other words, the lapse on the horizon serves as the volume flux density.

\section{Stationarity of maximal foliation and volume flow in the late-time regime}\label{App:FluxEquality}

In this appendix, we focus on the AdS-Schwarzschild in 3+1 dimensions, and explicitly check 
that the maximal foliation and volume flow are stationary at late times. 
To do this, we first write the Schwarzschild-AdS black hole in the ``maximal slicing'' gauge:
\begin{equation}\label{ADMmetric}
    ds^{2} = -\alpha^{2} d\tau^{2} + \gamma^2(dr+ \beta d\t)^2+ r^{2} d\Omega_{2}^{2},
\end{equation}
in which the 
slices of constant $\tau$ are the left-right symmetric maximal 
slices that asymptote to constant Schwarzschild time 
slices at the two boundaries. The radial coordinate $r$ and the angles $\theta$ and $\phi$ on the sphere can be chosen to be the same as the usual Schwarzschild coordinates. The functions $\alpha$, $\beta$ and $\gamma$ are functions of $\tau$ and $r$, given by:
\begin{equation}\label{gamma}
    \g^{-2} = 1 - \frac{2M}{r} + \frac{C{(\tau)}^{2}}{r^{4}} - \frac{\Lambda}{3} r^{2}
\end{equation}
\begin{equation}\label{alpha}
    \alpha = \gamma^{-1} \left[ 1 + C_{,\tau} \int_{r}^{\infty} \frac{\gamma^{3}{(\tau,r')}}{r'^{2}} dr' \right]
\end{equation}
\begin{equation}\label{beta}
    \beta = \frac{\alpha C{(\tau)}}{r^{2}}
\end{equation}
for some function $C{(\tau)}$. To derive equations (\ref{gamma}), (\ref{alpha}) and (\ref{beta}), we can, for example, feed the metric (\ref{ADMmetric}) into Einstein's equation. 
Analogous calculations for black holes in flat space have been done in the numerical relativity literature (see for example \cite{Estabrook}). 
In writing \eqref{alpha} we have imposed the boundary condition $\a\g\rightarrow1$ as $r\rightarrow\infty$, so that $\tau$ will agree asymptotically with the standard AdS global time coordinate.

The function $C{(\tau)}$ is determined by regularity of the maximal  
slice at the ``middle''.
According to \eqref{ADMmetric}, the metric induced on the maximal slice is 
$\g^2 dr^2 + r^2 d\O_2^2$, so in particular $\g dr$ is a unit 1-form on the slice. Since $r$ reaches a minimum 
at the middle of each slice, the pullback of $dr$ to the slice vanishes at the middle, so $\g$ must diverge there.
This  implies
\beq
    C^{2}=  2M r_{m}^{3} - r_{m}^{4} + \frac{\Lambda}{3} r_{m}^{6},
\eeq
where $r_m=r_m(\tau)$ is the $r$ coordinate at the middle of each constant $\tau$ slice. 
In the late time limit, $r_m(\t)$ approaches a constant, namely $r_f$, the radial coordinate of the
``final slice". Therefore $C(\t)$ too approaches a constant. It follows that the metric functions
$\a$, $\b$, and $\g$ all become constant in the late $\t$ limit, which  implies that the coordinate 
vector field $\partial_\tau$ approaches 
the Schwarzschild time Killing field. The unit normal 1-form
$\a d\t$ therefore 
becomes invariant under the Killing flow, as does the 
volume current 
\begin{equation}
    v = \alpha^{-1}(\partial_{\tau} - {\beta}\partial_{r}),
\end{equation}
which is minus the contravariant form of $\a d\t$. 

\section{Maximal 
slices in Vaidya: a closer look}\label{App:Vaidya}

Consider a maximal 
slice anchored at boundary time $t_{b}$
{in the double shell Vaidya/AdS spacetime in $2+1$ dimensions, \eqref{BTZVaidya}, \eqref{dubbelstep}).}
 For $t_{b} < 0$, the 
 slice stays entirely inside the AdS part of the geometry and is given by a constant $t$ slice (where $t$ denotes the global time in AdS). When $0 < t_{b} < b$, the 
 slice crosses the first shell and has two portions, one in the AdS region (still a constant $t$ slice) and one outside the shell. The volume functional for the part outside the shell reads:
\begin{equation}
    \mathrm{Vol} = 2\pi \int r\sqrt{2r' - f} dv
\end{equation}
where we write the 
slice as a function $r(v)$. Since the functional is independent of $v$, we have a conserved energy:
\begin{equation}
    E = r' \frac{\partial L}{\partial r'} - L = \frac{r(f-r')}{\sqrt{2r'-f}}
\end{equation}
Similarly, there is a conserved energy in the AdS region, which can be shown to vanish by smoothness at the center ($r=0$).
The maximal 
slices consist of locally maximal 
slices apart from on the shells, where they satisfy a Weierstrass-Erdmann corner condition \cite{gelfand2012calculus}. Since the shells are located at a constant value of $v$, the corner 
condition simplifies to the requirement that the ``conjugate momentum"
\begin{equation}
    p_{r} = \frac{\partial L}{\partial r'} = \frac{r}{\sqrt{2r' - f}}
\end{equation}
be continuous across the junction, which amounts to requiring that the jump in $r'$ is 1/2 the jump in $f$. Together with the fact that the portion of the maximal 
slice in the AdS region must be constant global time slice, 
this determines the derivative of $r(v)$ at the junction on the BTZ side:
\begin{equation}
    \frac{dr}{dv} \bigg|_{r_{1,+}} = 1 + r_{1}^{2} - \frac{a}{2}(1+r_{+}^{2})
\end{equation}
where $r_{1}$ is the $r$-coordinate of this junction, and $r_{1,+}$ means an $r$ value slightly larger.
Similarly, for $t_{b} > b$, the maximal slice crosses both shells and the junction condition %above 
has to be imposed at each junction. At the outer junction, located at $r=r_{2}$, 
the discontinuity of the derivative of $r(v)$ is found to be:
\begin{equation}
    \frac{dr}{dv} \bigg|_{r_{2,-}} - \frac{dr}{dv} \bigg|_{r_{2,+}} = \frac{1-a}{2}(1+r_{+}^{2}).
\end{equation}

\bibliographystyle{JHEP}
\bibliography{ThreadsBib.bib} 

\end{document}